\documentclass{aa}

\usepackage{graphicx}
\usepackage{amsmath,amssymb,bm}
\usepackage{booktabs}
\usepackage{tikz}
\usetikzlibrary{arrows.meta,fit}
\usepackage{orcidlink}
\usepackage{CJKutf8}

\newcommand{\msun}{\mathrm{M}_{\odot}}
\newcommand{\MG}{M_{\mathrm{G}}}
\newcommand{\Mtot}{m_{\rm tot}}
\newcommand{\mh}{[\mathrm{M}/\mathrm{H}]}
\newcommand{\dd}{\,\mathrm{d}}
\newcommand{\ut}{\widetilde u}
\newcommand{\zh}{\widehat z}
\newcommand{\StudentT}{\mathcal{T}}
\newcommand{\Normal}{\mathcal{N}}
\newcommand{\uunit}{\mathrm{km\,s^{-1}\,au^{1/2}}}
\newcommand{\cn}[1]{\begin{CJK*}{UTF8}{gbsn}#1\end{CJK*}}

\newcommand{\softplus}{\operatorname{softplus}}
\newcommand{\sigmoid}{\operatorname{sigmoid}}

\begin{document}

\title{Dynamical calibration of a metallicity-dependent stellar mass--luminosity relation from Gaia wide binaries}
\author{Yu-Ting Wang (\cn{王雨亭})\inst{1}\orcidlink{0009-0007-8574-0890}\email{yuwang@mpia.de}
        \and Hans-Walter Rix\inst{1}\orcidlink{0000-0003-4996-9069}\email{rix@mpia.de}
        \and Yuan-Sen Ting (\cn{丁源森})\inst{2,3,1,5}\orcidlink{0000-0001-5082-9536}\email{ting.74@osu.edu}
        \and Jiadong Li (\cn{李佳东})\inst{1}\orcidlink{0000-0002-3651-5482}\corrauth{jdli@mpia.de}
        \and Chao Liu (\cn{刘超})\inst{4}\orcidlink{0000-0002-1802-6917}\email{liuchao@nao.cas.cn}
        }

\institute{Max-Planck-Institut f\"ur Astronomie, K\"onigstuhl 17, 69117 Heidelberg, Germany
          \and Department of Astronomy, The Ohio State University, 140 West 18th Avenue, Columbus, OH 43210, USA
          \and Center for Cosmology and AstroParticle Physics (CCAPP), The Ohio State University, Columbus, OH 43210, USA
          \and National Astronomical Observatories, Chinese Academy of Sciences, 20A Datun Road, Chaoyang District, Beijing 100101, China
          \and Asia School of Business, No. 1 \& 11, Jalan Dato Onn, 50480 Kuala Lumpur, Malaysia
          }

\abstract{The mass of a main-sequence star is usually estimated from its luminosity through a mass--luminosity relation (MLR). This relation depends on metallicity, and its empirical calibration relies on few stars with measured masses. We calibrate the MLR of single main-sequence stars in the Gaia $G$ band as a function of metallicity with $10\,559$ wide binaries. The projected relative speed and separation of a wide binary constrain its total mass statistically. We describe the stellar mass as a monotone function of absolute magnitude $\MG$ and metallicity $\mh$ and fit this function to all pairs. The metallicities come from Gaia low-resolution (XP) spectra and are calibrated by requiring that both stars of a pair have the same metallicity. The likelihood includes the measurement noise, the selection of the sample and pairs with an undetected close companion. The data constrain the MLR over $3.7\lesssim\MG\lesssim11.8$\,mag and $-0.5\lesssim\mh\lesssim0.4$ on our calibrated metallicity scale, with a median statistical uncertainty of $2.5\%$ in mass. Tests of the model assumptions indicate an additional systematic uncertainty of up to $5\%$. At solar metallicity the MLR agrees with the PARSEC models to within $5\%$. At the faint end, the MIST, BaSTI, Dartmouth and BHAC15 models predict masses 8 to $22\%$ below our MLR. At fixed $\MG$ the mass increases with metallicity, by $12\%$ per $0.5$\,dex at $\MG=5$\,mag and by $26\%$ at $\MG=11$\,mag. The dependence is $1\sigma$ to $2\sigma$ stronger than PARSEC predicts. Published masses from orbits, eclipsing binaries and asteroseismology lie within about $10\%$ of our MLR for $\MG<10.5$\,mag. For fainter stars they are about $17\%$ lower. The origin of the difference at the faint end is not yet understood. Because a shift of $0.1$\,dex in metallicity changes the mass by up to $5\%$, MLRs should be compared on a common metallicity scale. The MLR is available as the Python package \texttt{wbmlr} (\protect\url{https://github.com/yt-wangg/wide-binary-mlr}).}


\titlerunning{Dynamical calibration of the stellar MLR}
\authorrunning{Wang et al.}

\maketitle
\nolinenumbers

\section{Introduction}\label{sec:intro}

Stars less massive than the Sun dominate the stellar content of the Galaxy. M dwarfs account for about three quarters of the stars within 10\,pc \citep{Henry2018}. Stars below $1\,\msun$ also contain a large fraction of the stellar mass of the Galactic disc \citep{Chabrier2001}. 
The initial mass function inferred from star counts, and every quantity derived from it, such as the mass-to-light ratio of stellar populations, the contribution of dwarfs to the dynamical mass budget of the disc and the claimed variations of the initial mass function with environment and metallicity, depends directly on how masses are assigned to dwarfs. Low-mass stars are also the most common hosts of small planets \citep{Dressing2015}, so the masses of these stars also set the inferred properties of their planets. Stellar mass is nevertheless rarely measured directly, and for most stars it is inferred from the luminosity through a mass--luminosity relation (MLR).

An empirical MLR must be calibrated on stars with measured masses. Double-lined eclipsing binaries provide the masses of both stars to a few per cent \citep{Torres2010}. Visual and astrometric orbits provide masses for nearby binaries monitored over years to decades \citep{Henry1993,Delfosse2000,Benedict2016,Mann2019}. Spectroscopic binaries without eclipses give the projected orbit only. For double-lined systems (SB2) the radial velocities of both stars yield $m\sin^3 i$ for each component, and for single-lined systems (SB1) only the mass function \citep{Shahaf2019}. Gaia astrometric orbits, combined with spectroscopic orbits, now provide masses for many more such binaries \citep{GaiaCollaboration2023NSS,Chevalier2023}. Asteroseismology provides masses of single stars, although the oscillations of K and M dwarfs are too weak to detect \citep{ChaplinMiglio2013,Serenelli2017,Pinsonneault2018}.

The calibrating samples remain small, and the most precise relation for M dwarfs is based on 62 binaries \citep{Mann2019}. That relation gives masses to between 2 and $3\%$ in the $K_s$ band and shows no measurable dependence on metallicity near the solar value. Optical bands such as Gaia $G$ are more sensitive to metallicity because of molecular absorption, and there the dependence is poorly constrained by observations. Stellar models cannot yet replace an empirical calibration, because their optical magnitudes for M dwarfs remain uncertain \citep{Baraffe2015,Chen2014}.

\begin{table}
\caption{Selection of the sample.}
\label{tab:cuts}
\centering
\begin{tabular}{@{}ll@{}}
\toprule
Quantity & Cut \\
\midrule
\texttt{R\_chance\_align} & $\le0.01$ \\
Projected separation $s$ & $<10^4$\,au \\
Parallax & $>1$\,mas \\
Astrometric fidelity & $\ge0.5$ \\
Gaia G/BP/RP-band flux S/N & $>10$ \\
Corrected BP/RP flux excess $|C^{*}|$ & $\le3\sigma_{C^{*}}$ \\
Fraction of blended BP/RP transits & $<50\%$ \\
Extinction $A_G$, primary & $\le0.1$\,mag \\
J-CaPS fit & converged \\
J-CaPS parameters & not at a bound \\
J-CaPS uncertainties & flagged reliable \\
J-CaPS parallax & flagged reliable \\
J-CaPS extinction, primary & flagged reliable \\
Single-star criterion, $\Delta\mathrm{BIC}$ & $>0$ \\
$\MG$ & $3.5$ to $13.5$\,mag \\
$u/\sigma_u$ & $>3$ \\
\bottomrule
\end{tabular}

\tablefoot{The J-CaPS cuts, the single-star criterion and the $\MG$ range apply to both stars of a pair unless the primary is named. The astrometric fidelity is that of \citet{Rybizki2022}, and the two BP/RP cuts follow \citet{Riello2021GaiaEDR3Photometry}.}
\end{table}

Wide binaries allow a calibration with far larger samples. Their orbital periods of $10^4$ to $10^6$\,yr, at typical separations of $10^3$ to $10^4$\,au, are too long for an orbit to be followed. A wide binary is observed at a single snapshot, which gives the projected separation $s$ and the projected relative speed $v$ of the two components. Kepler's law nevertheless connects these two quantities to the total mass $\Mtot$, because $v\sqrt{s}$ is proportional to $\sqrt{\Mtot}$. The constant of proportionality depends on the phase, eccentricity and orientation of the orbit, which are unknown for an individual pair. For a population of pairs its distribution can be computed, so that the masses can be inferred statistically. 

Gaia made this approach practical by providing astrometry for about a million candidate wide binaries \citep{ElBadry2021MillionBinaries}. An MLR in the $G_{\rm RP}$ band was derived in this way from a few thousand pairs \citep{Giovinazzi2022}. The method was then extended to the whole colour--magnitude diagram by \citet{Hwang2024DynamicalMasses}. Their neural network gives a mean stellar mass for each position in the diagram.

As an MLR of single stars, this map is limited in several respects. Because it gives the mean mass of all objects at a given position, it averages single stars with the unresolved binaries that overlap them. Metallicity is absent from the map, and the selection of the sample is absent from its likelihood. Its uncertainties are the scatter of repeated network predictions, not a posterior that includes the measurement errors, a model of unresolved companions and the sample selection. 

In this paper we calibrate the MLR of single main-sequence stars as a function of the absolute $G$ magnitude $\MG$ and metallicity. We use the pairs of \citet{ElBadry2021MillionBinaries} for which both stars have a metallicity from Gaia low-resolution BP/RP (XP) spectra \citep{Li2026JCaPS}. Pairs in which an XP spectrum shows a companion are removed, and pairs in which a fainter companion goes undetected are modelled explicitly. The likelihood also includes the measurement noise and the selection of the sample. The resulting MLR is defined for $3.5\le\MG\le13.5$\,mag, which corresponds to masses from about $0.1\,\msun$ to more than $1\,\msun$.

The relation is available as a table, a set of posterior draws and a small Python package, \texttt{wbmlr} (\url{https://github.com/yt-wangg/wide-binary-mlr}; see Sect.~\ref{sec:release}), so that it can be applied to other samples. The paper is organised as follows. Section~\ref{sec:data} describes the data, Sect.~\ref{sec:method} the inference, and Sect.~\ref{sec:results} the resulting MLR. In Sect.~\ref{sec:discussion} we compare the relation with individual mass measurements and with stellar models, and discuss the prospects with Gaia DR4 and the limitations of this work. We conclude in Sect.~\ref{sec:conclusions}. The appendices give the full model specification (Appendix~\ref{app:model}), the forward model of hidden companions that sets the outlier component (Appendix~\ref{app:companions}), the posterior computation (Appendix~\ref{app:inference}) and the tests of the main modelling assumptions (Appendix~\ref{sec:systematics}).


\begin{figure}[t]
\centering
\includegraphics[width=\columnwidth]{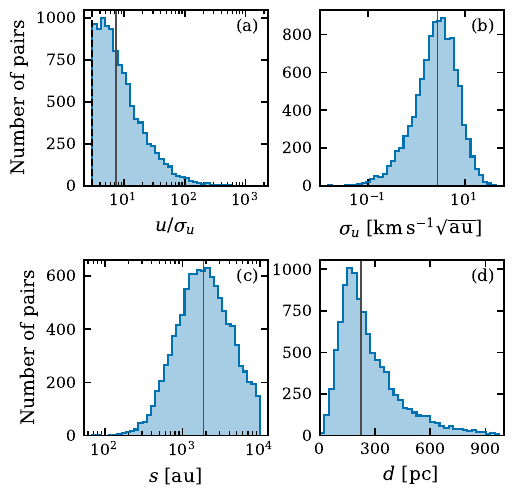}
\caption{Distributions of $u/\sigma_u$ (a), $\sigma_u$ (b), projected separation $s$ (c) and distance $d$ (d) for the $10\,559$ pairs of the sample. Solid lines mark the medians, and the dashed line in panel (a) marks the cut at $u/\sigma_u=3$.}
\label{fig:sample}
\end{figure}

\begin{figure*}[t]
\centering
\begin{tikzpicture}[
    x=1cm,y=1cm,
  var/.style={circle,draw,minimum size=12mm,inner sep=1pt,font=\normalsize},
  obs/.style={var,fill=black!16},
  det/.style={var,double,double distance=1.2pt},
  par/.style={draw,rounded corners=6mm,minimum height=12mm,minimum width=12mm,inner xsep=7pt,inner ysep=0pt,font=\normalsize},
  dot/.style={circle,fill,inner sep=1.8pt},
  arr/.style={-{Latex[length=2.4mm]},semithick},
  plate/.style={draw,rounded corners=3pt},
  lab/.style={font=\normalsize,inner sep=2pt}
]
\node[par] (b)  at (1.6,4.9) {$b,\,s_z,\,\nu$};
\node[par] (om) at (4.4,4.9) {$\bm\omega$};
\node[par] (th) at (7.2,4.9) {$\Theta$};
\node[par] (fb) at (12.4,4.9) {$f_0,\,\beta$};
\node[lab] at (3.0,5.95) {metallicity (Sect.~\ref{sec:zcal})};
\node[lab] at (7.2,5.95) {MLR (Sect.~\ref{sec:mlr})};
\node[lab] at (11.5,5.95) {likelihood (Sects.~\ref{sec:dynamics} and \ref{sec:selection})};
\node[obs] (zh) at (1.6,2.6) {$\zh_{jk}$};
\node[var] (Z)  at (4.4,2.6) {$Z_j$};
\node[obs] (MG) at (1.6,0.5) {$M_{\mathrm{G},jk}$};
\node[det] (ka) at (7.2,0.5) {$\kappa_j$};
\node[det] (f)  at (12.4,2.6) {$f_j$};
\node[var] (ut) at (12.4,0.5) {$\ut_j$};
\node[det] (u0) at (9.8,0.5) {$u_{0,j}$};
\node[obs] (u)  at (9.8,-1.7) {$u_j$};
\node[lab,anchor=east] (sel) at (8.6,-1.7) {in the sample if $3\sigma_{u,j}<u_j<U_j$};
\node[dot] (sz) at (-0.5,2.6) {};
\node[lab,anchor=south] (szl) at (-0.5,2.75) {$\sigma_{z,jk}$};
\node[dot] (rj) at (14.4,2.6) {};
\node[lab,anchor=south] (rjl) at (14.4,2.75) {$r_j$};
\node[dot] (hj) at (14.4,0.5) {};
\node[lab,anchor=south] (hjl) at (14.4,0.65) {$p,\,h_j$};
\node[dot] (su) at (11.5,-1.7) {};
\node[lab,anchor=west] (sul) at (11.65,-1.7) {$\sigma_{u,j},\,U_j$};
\node[plate,fit=(zh)(MG)(sz)(szl),inner sep=9pt] (pk) {};
\node[lab,anchor=north west] at ([yshift=-2pt]pk.south west) {stars $k=1,2$};
\node[plate,fit=(pk)(Z)(ka)(f)(ut)(u0)(u)(sel)(rj)(rjl)(hj)(hjl)(su)(sul),inner xsep=11pt,inner ysep=14pt] (pj) {};
\node[lab,anchor=south east] at ([xshift=-5pt,yshift=3pt]pj.south east) {pairs $j$};
\draw[arr] (b) -- (zh);
\draw[arr] (sz) -- (zh);
\draw[arr] (om) -- (Z);
\draw[arr] (Z) -- (zh);
\draw[arr] (MG) -- (zh);
\draw[arr] (MG) -- (ka);
\draw[arr] (Z) -- (ka);
\draw[arr] (th) -- (ka);
\draw[arr] (fb) -- (f);
\draw[arr] (rj) -- (f);
\draw[arr] (f) -- (ut);
\draw[arr] (hj) -- (ut);
\draw[arr] (ka) -- (u0);
\draw[arr] (ut) -- (u0);
\draw[arr] (u0) -- (u);
\draw[arr] (su) -- (u);
\end{tikzpicture}
\caption{Graphical model of the fit. Shaded circles are observed quantities, open circles are unknown quantities, double circles are derived quantities, and dots are fixed inputs. On the left, Eqs.~(\ref{eq:zmodel}) and (\ref{eq:zpost}) give the metallicity $Z_j$ of pair $j$, which the MLR converts into the mass factor $\kappa_j$ (Eqs.~(\ref{eq:g}) and (\ref{eq:kappa})). On the right, the orbit factor $\ut_j$ follows $p(\ut)$ or, with probability $f_j$, the outlier distribution $h_j$ (Eqs.~(\ref{eq:mix}) and (\ref{eq:fout})). We observe the noise-free value $u_{0,j}=\kappa_j\ut_j$ as $u_j$, with measurement noise and within the selection limits (Eqs.~(\ref{eq:rice}) to (\ref{eq:sel})). Parameters in the top row describe the whole sample, and those of the metallicity part are fitted first and then held fixed.}
\label{fig:pgm}
\end{figure*}
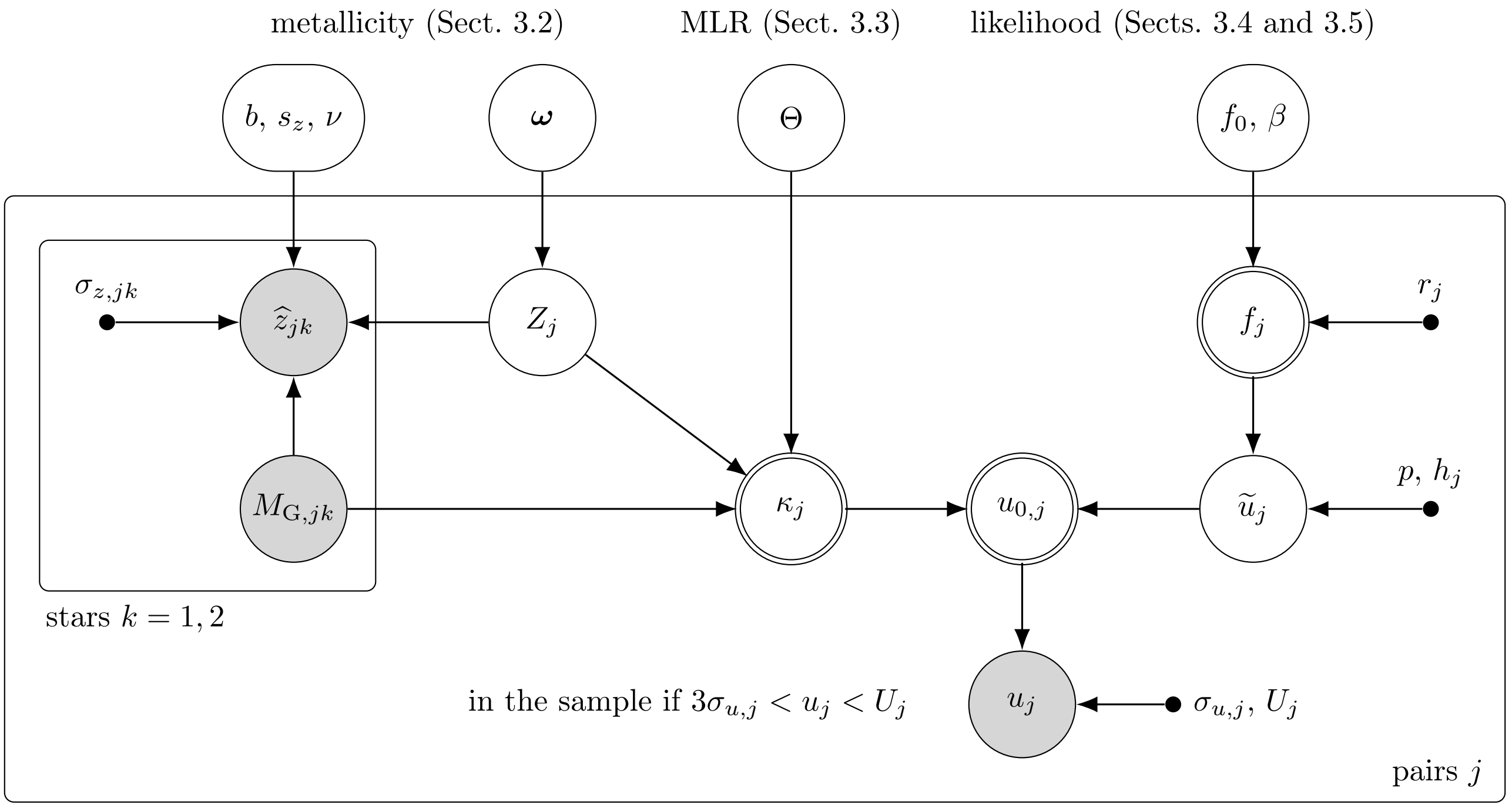

\section{Data}\label{sec:data}
\subsection{Wide-binary sample}\label{sec:cat}
We use the catalogue of \citet{ElBadry2021MillionBinaries}, which contains about $1.3\times10^6$ candidate wide binaries within about 1\,kpc, selected from Gaia EDR3 \citep{GaiaCollaboration2021EDR3}. A pair enters this catalogue only if its proper-motion difference is consistent with a bound orbit. The catalogue also provides an estimate of the probability that a pair is a chance alignment (\texttt{R\_chance\_align}). Our cuts, which are listed in Table~\ref{tab:cuts}, select pairs with a low probability of chance alignment, reliable astrometry, photometry and J-CaPS fits, and low extinction.

Both stars of a pair are assigned the same distance $d$, which we derive from the parallax of the primary (the brighter star). The parallax is corrected for the zero point of \citet{Lindegren2021ZP}, and the extinction $A_G$ is taken from the dust map of \citet{Edenhofer2024}. The absolute magnitude of each star is then $\MG=G+5-5\log_{10}(d/\mathrm{pc})-A_G$.

\subsection{Metallicities and removal of unresolved binaries}\label{sec:jcaps}
Our metallicities come from J-CaPS \citep{Li2026JCaPS}, a code that fits a neural-network model to each Gaia XP spectrum \citep{GaiaCollaboration2023DR3}. The parameters of the model are the effective temperature, the absolute magnitude and the metallicity of the star. From this fit we use the metallicity, whereas all absolute magnitudes in this paper are the photometric values of Sect.~\ref{sec:cat}. We denote the J-CaPS metallicity $\mh$ of a star by $\zh$ and its uncertainty by $\sigma_z$. These raw metallicities are least certain for metal-poor stars and for cool dwarfs, where the training set of J-CaPS is sparse. They also have a bias that depends on the absolute magnitude of the star, which we calibrate in Sect.~\ref{sec:zcal}.

The single-star criterion is taken from the analysis of Gaia XP spectra of \citet{Li2025BinaryXP}, who fit each spectrum with a single-star model and with a binary model that adds a companion described by its mass ratio. We keep the pairs whose components are both classified as single, that is, for which the Bayesian information criterion favours the single-star model, $\Delta\mathrm{BIC}=\mathrm{BIC}_{\rm binary}-\mathrm{BIC}_{\rm single}>0$. This removes the unresolved binaries whose companion contributes detectably to the XP spectrum. Binaries with fainter companions remain in the sample, and are described by the outlier component of the dynamical likelihood, whose shape we compute from a forward model simulation of such companions (Sect.~\ref{sec:dynamics} and Appendix~\ref{app:companions}).

\begin{figure*}[t]
\centering
\includegraphics[width=0.82\textwidth]{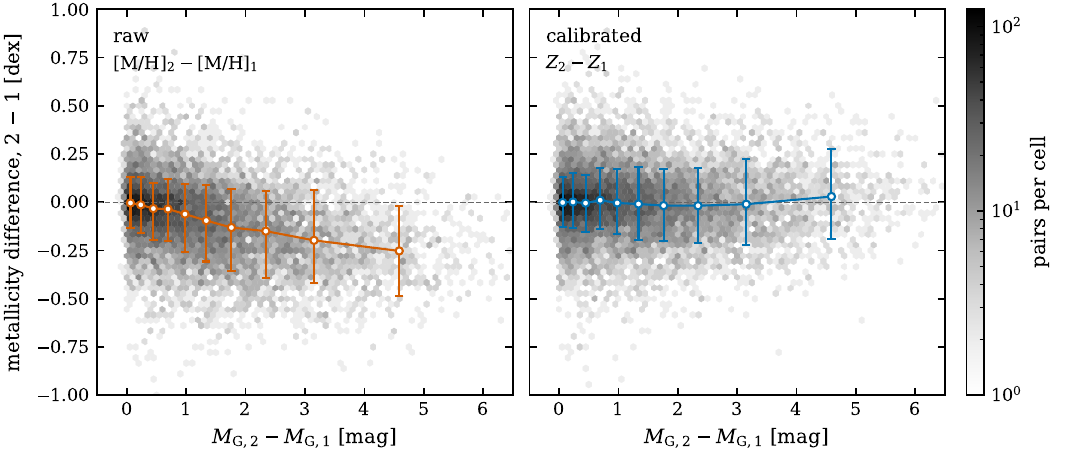}
\caption{Metallicity difference of the two stars of a pair against their difference in absolute magnitude, before (left) and after (right) the calibration. Star 1 is the primary. Grey shading gives the number of pairs. Points are medians in ten bins with equal numbers of pairs, and bars mark the central $68\%$ of each bin.}
\label{fig:zdelta}
\end{figure*}

\subsection{Scaled relative velocity}\label{sec:velocity}
The dynamical information of a pair is contained in the angular separation of its two stars and in their proper-motion difference $\Delta\mu$. Together with the distance $d$, these quantities give the projected separation $s$ in au and the projected relative speed $v$ in $\mathrm{km\,s^{-1}}$. As in \citet{Hwang2024DynamicalMasses}, we combine the two into the scaled velocity
\begin{equation}
 u=v\sqrt{s},
 \label{eq:u}
\end{equation}
which we express in units of $\uunit$ throughout. It is a measure of the total mass of the pair (Sect.~\ref{sec:overview}).

The uncertainty of $u$ is dominated by that of the proper-motion difference, so we give $u$ the same relative uncertainty,
\begin{equation}
 \sigma_{u}=\frac{u}{\Delta\mu/\sigma_{\Delta\mu}},
 \label{eq:sigu}
\end{equation}
where $\sigma_{\Delta\mu}$ is the uncertainty of $\Delta\mu$ along the direction of the proper-motion difference. Because Gaia underestimates the astrometric uncertainties of bright stars, we first multiply the proper-motion uncertainties by the inflation factor of \citet[][their Eq.~16]{ElBadry2021MillionBinaries}. This factor was measured for parallaxes, which come from the same astrometric solution, and it reaches $1.3$ near $G=12.7$\,mag. For our pairs it increases $\sigma_u$ by $5\%$ in the median. We treat the errors of the two stars as independent and neglect the parallax uncertainty, which is small in comparison.

As a final cut we require that the proper-motion difference is detected at more than $3\sigma$, which is equivalent to $u/\sigma_u>3$. The resulting sample contains $10\,559$ pairs. Figure~\ref{fig:sample} shows their distributions of $u/\sigma_u$, $\sigma_u$, separation and distance. For most pairs the proper-motion difference is detected at moderate significance, with a median $u/\sigma_u$ of $7.3$.

The sample is truncated in $u$ on both sides. Our cut removes low-mass pairs at large distances and separations, for which the proper-motion difference is small. The bound-orbit criterion of the catalogue, in turn, removes pairs with large $u$. Both truncations are included in the model (Sect.~\ref{sec:selection}).

\section{Method}\label{sec:method}
\subsection{Overview}\label{sec:overview}
The orbit factor $\ut$, which relates $u$ to the total mass of a pair (Sect.~\ref{sec:dynamics}), is unknown for an individual pair, but its distribution $p(\ut)$ over many pairs can be computed. A larger total mass moves the whole distribution of the noise-free $u$ to larger values. The scaled velocities of a group of similar pairs then determine their total mass. An MLR allows all pairs to be fitted simultaneously, without the need to group them. For given parameters, the MLR predicts the mass of each star from its $\MG$ and metallicity, and hence the total mass of each pair. The parameters are adjusted until the predicted distributions of $u$ match the observed values for all pairs.

The model consists of three parts, which are summarised in Fig.~\ref{fig:pgm}. We first calibrate the XP metallicities, which provides a probability distribution for the metallicity of each pair (Sect.~\ref{sec:zcal}), and hold this result fixed in the dynamical fit. The MLR is a smooth surface in $\MG$ and metallicity (Sect.~\ref{sec:mlr}). The likelihood of $u$ includes hidden companions, measurement noise and the selection of the sample (Sects.~\ref{sec:dynamics} and \ref{sec:selection}). Throughout, the index $j$ denotes a pair and $k$ its two stars, with $k=1$ for the primary, whereas symbols without $j$ describe the whole sample.

\subsection{Metallicity calibration}\label{sec:zcal}
Each pair requires a metallicity for the dynamical fit, and the XP metallicities have a bias as well as large noise. We remove the bias with the pairs themselves and reduce the noise with a prior. The two stars of a wide binary formed from the same gas, and their abundances agree to high precision in most cases \citep{Hawkins2020}. We therefore assume that both stars of pair $j$ have the same metallicity $Z_j$. A raw measurement $\zh_{jk}$ then differs from $Z_j$ by noise and by a bias $b$ that depends on its absolute magnitude,
\begin{equation}
 \zh_{jk}\sim\StudentT_{\nu}\!\left[Z_j+b(M_{\mathrm{G},jk}),\ \sqrt{\sigma_{z,jk}^2+s_z^2}\,\right],
 \label{eq:zmodel}
\end{equation}
where $\StudentT_{\nu}[\mu,\sigma]$ denotes a Student-$t$ distribution with $\nu$ degrees of freedom, location $\mu$ and scale $\sigma$, and $s_z$ is an additional scatter. We use a Student-$t$ distribution because its broad wings allow for occasional large errors.

Because $Z_j$ cancels in the difference of the two measurements of a pair, these differences determine the bias. We approximate their distribution by a Student-$t$ distribution with the same $\nu$,
\begin{equation}
 \zh_{j1}-\zh_{j2}\sim\StudentT_{\nu}\!\bigl[b(M_{\mathrm{G},j1})-b(M_{\mathrm{G},j2}),\,(\sigma_{z,j1}^2+\sigma_{z,j2}^2+2s_z^2)^{1/2}\bigr],
 \label{eq:zdiff}
\end{equation}
and fit a piecewise-linear $b$, together with $s_z$ and $\nu$, by maximum likelihood.

\begin{figure*}[t]
\centering
\includegraphics[width=0.85\textwidth]{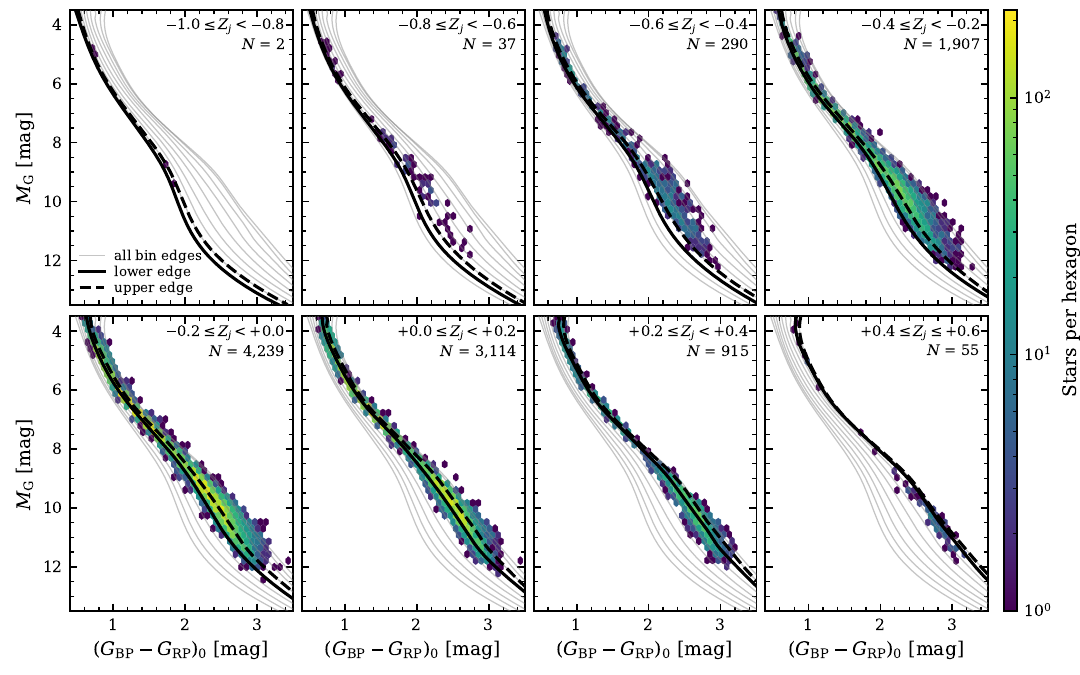}
\caption{Colour--magnitude diagrams of the stars of the sample in bins of the metallicity of their pair (the median of $p_j(Z)$). Black solid and dashed lines are the PARSEC main sequences at the lower and upper edge of each bin, for an age of 4\,Gyr. Grey lines repeat the main sequences at all bin edges. The main sequence moves to redder colours with increasing metallicity.}
\label{fig:cmd}
\end{figure*}

Figure~\ref{fig:zdelta} shows the differences before and after the calibration. The raw metallicities are too low for faint stars, by $0.25$\,dex for two stars that differ by $4.7$\,mag. The calibration removes this trend to within $0.03$\,dex. This agreement is partly by construction, since $b$ was fitted to the same differences. The figure nevertheless shows that a bias that depends on $\MG$ is sufficient to remove the trend. Any real metallicity difference between stars of different mass, for example from atomic diffusion \citep{Dotter2017Diffusion}, would be included in $b$.

The differences within pairs do not constrain the zero point of the metallicity scale. We fix it by assuming that J-CaPS is unbiased at the absolute magnitude of the Sun (4.67\,mag), where we set $b=0$. We refer to $\zh-b(\MG)$ as the calibrated metallicity of a star and use the symbol $Z$ for metallicities on this calibrated scale. Like $\mh$, $Z$ is logarithmic and equals zero for the Sun. It is not the metal mass fraction.

The two measurements of a pair define a likelihood for $Z_j$ through Eq.~(\ref{eq:zmodel}). This likelihood is broad, since the central $68\%$ of the differences within pairs is $0.3$ to $0.5$\,dex wide (Fig.~\ref{fig:zdelta}). We therefore combine it with a prior $p_{\rm pop}(Z)$ that describes the metallicity distribution of the whole sample,
\begin{equation}
   p_j(Z)\propto p_{\rm pop}(Z)\,p(\zh_{j1}\mid Z)\,p(\zh_{j2}\mid Z),
 \label{eq:zpost}
\end{equation}
where $p(\zh_{jk}\mid Z)$ is the density of Eq.~(\ref{eq:zmodel}) at the measured value. With this prior, metallicities that are common in the sample become more probable for a pair with noisy measurements. We represent $p_{\rm pop}$ by a sum of six broad Gaussians and fit their fractions $\bm\omega$ to all pairs of the sample together (Appendix~\ref{app:model}). The resulting $p_j(Z)$ is the probability distribution of the metallicity of pair $j$. Figure~\ref{fig:cmd} shows the colour--magnitude diagrams of the sample in bins of the calibrated metallicity $Z$, taken as the median of $p_j(Z)$ of each pair, and compares them with PARSEC isochrones at the edges of each bin.

\subsection{Parameterisation of the MLR}\label{sec:mlr}
The MLR should follow the data where they are dense and remain smooth where they are sparse. We describe it by a spline surface that is constrained by two physical conditions and by priors centred on a stellar model. The surface gives the stellar mass $m$ as a function of absolute magnitude and calibrated metallicity,
\begin{equation}
 \log_{10}\frac{m(\MG,Z)}{\msun}=\sum_{i=1}^{8}\sum_{l=1}^{4}\Theta_{il}\,\phi_i(\MG)\,\psi_l(Z),
 \label{eq:g}
\end{equation}
where $\phi_i$ and $\psi_l$ are cubic B-splines \citep{deBoor2001Splines} and $\Theta_{il}$ are the coefficients. The surface is defined for $3.5\le\MG\le13.5$\,mag and $-1.0\le Z\le0.6$. Its bright limit excludes giants, whose mass at a given $\MG$ depends on age.

We require that the mass is a non-increasing function of $\MG$ at fixed metallicity and a non-decreasing function of metallicity at fixed $\MG$. For the spline coefficients these two conditions read
\begin{equation}
 \Theta_{i+1,l}\le\Theta_{il},\qquad \Theta_{i,l+1}\ge\Theta_{il}.
 \label{eq:mono}
\end{equation}
The conditions fix the signs of the two slopes of the surface, whereas the sizes of the slopes are determined by the data.

We write the coefficients as $\Theta=\Pi+D$, where $\Pi$ represents the PARSEC relation at an age of 4\,Gyr \citep{Bressan2012PARSEC} and $D$ is a fitted correction. A shrinkage prior keeps $D$ small, and a smoothness prior keeps it smooth in regions with few data (Appendix~\ref{app:model}). The shrinkage prior has a standard deviation of $0.10$\,dex, or $26\%$ in mass. This is several times the statistical uncertainty of the MLR where the data are dense, which is 2 to $5\%$ (Sect.~\ref{sec:res_mlr}). The data thus mainly determine the MLR there.

The conditions of Eq.~(\ref{eq:mono}) apply to the total surface, so the correction $D$ can have either sign. They do not restrict how the metallicity dependence varies with $\MG$. That variation is regularised by the smoothness prior, which favours the variation of PARSEC where the data are sparse.

A final prior constrains the mass scale at the solar point, which we define by the $\MG$ and metallicity of the Sun. This prior, the solar mass anchor, gives stars at the solar point a mean mass of $1\,\msun$ with a standard deviation of $5\%$. The width allows for the range of ages of such stars. On MIST isochrones their mass decreases from $1.03\,\msun$ at 1\,Gyr to $0.91\,\msun$ at 10\,Gyr \citep{Choi2016MIST}. Its effect on the MLR is tested in Sect.~\ref{sec:tests}.

Colour is not included as a third variable. On the main sequence the $G_{\rm BP}-G_{\rm RP}$ colour at a given $\MG$ is determined mainly by metallicity (Fig.~\ref{fig:cmd}), so it would add little information. Metallicity is also the physical variable, and many spectroscopic and photometric surveys provide it. For users without a metallicity, the public tables give the mass as a function of colour and $\MG$ as well.

\begin{figure}[t]
\centering
\includegraphics[width=\columnwidth]{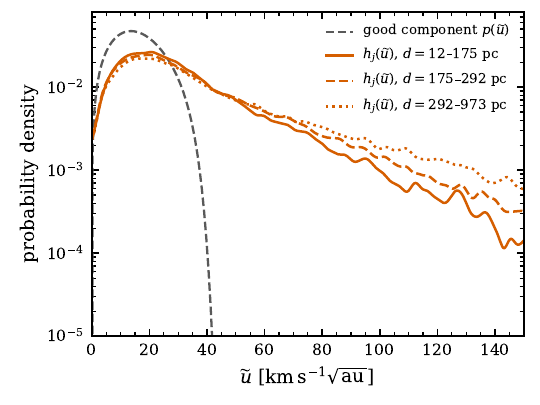}
\caption{Distribution of the orbit factor $\ut$ for the good component (grey dashed line) and for the outlier component (orange lines). The outlier distributions $h_j$ are averaged over the pairs in each third of the distance distribution. All curves are normalised to unit area.}
\label{fig:template_dist}
\end{figure}

\subsection{Dynamical likelihood}\label{sec:dynamics}
We build the dynamical likelihood on that of \citet{Hwang2024DynamicalMasses}, and extend it with a measurement model that holds at low signal-to-noise ratio, the metallicity dependence of the masses and the selection of the sample.

\paragraph{Orbit factor.} In a bound binary, the total mass $m_{\rm tot}=m_1+m_2$, the semi-major axis $a$, the eccentricity $e$ and the orbital phase set the instantaneous relative speed,
\begin{equation}
 v_{\rm rel}=\sqrt{\frac{Gm_{\rm tot}}{a}}\,f(\phi,e),
\end{equation}
where $f(\phi,e)$ describes the dependence on the orbital phase $\phi$ along the ellipse, from $f=\sqrt{(1+e)/(1-e)}$ at pericentre to $f=\sqrt{(1-e)/(1+e)}$ at apocentre. We observe only the projected speed $v$ and the projected separation $s$, and use the statistic $u=v\sqrt{s}$ of Eq.~(\ref{eq:u}). Because $v\propto a^{-1/2}$ and $s\propto a$, the semi-major axis cancels. The orbit factor $\ut=u_0/\sqrt{m_{\rm tot}/\msun}$, where $u_0$ is the value of $u$ without measurement noise, therefore depends only on the eccentricity, the phase and the orientation of the orbit. For reference, a circular orbit seen face-on has $\ut=29.8$ in our units, which is the orbital speed of the Earth in $\mathrm{km\,s^{-1}}$. Marginalising over these parameters in a simulation of orbits gives the distribution $p(\ut)$, and $u_0$ then has the density
\begin{equation}
 p(u_0\mid m_{\rm tot})=\frac{1}{\sqrt{m_{\rm tot}/\msun}}\,p\!\left(\frac{u_0}{\sqrt{m_{\rm tot}/\msun}}\right).
\end{equation}

\paragraph{From the MLR to the likelihood of a pair.} For a metallicity $Z$, the MLR of Eq.~(\ref{eq:g}) gives the mass $m_{jk}(Z)=m(M_{\mathrm{G},jk},Z)$ of star $k$ in pair $j$, so that the total mass of the pair is
\begin{equation}
 m_{{\rm tot},j}(Z)=m_{j1}(Z)+m_{j2}(Z),
\end{equation}
and its mass scale is
\begin{equation}
 \kappa_j(Z)=\sqrt{m_{{\rm tot},j}(Z)/\msun},
 \label{eq:kappa}
\end{equation}
so that $u_{0,j}=\kappa_j\ut$.

The parameters $\Theta$ of the MLR therefore enter the dynamical likelihood only through the total masses of the pairs, and the metallicity $Z$ of each system is a nuisance parameter that is marginalised over its posterior (Sect.~\ref{sec:selection}).

\paragraph{Measurement errors.} The proper-motion difference is a two-dimensional vector with independent Gaussian errors of the same scale in both coordinates, so its magnitude $\Delta\mu$ follows a Rice distribution \citep{Rice1945}. At fixed distance and separation $u\propto\Delta\mu$, so the observed $u_j$ follows the same distribution, given $u_0$ and the scale $\sigma_{u,j}$ of Eq.~(\ref{eq:sigu}),
\begin{equation}
 R(u_j\mid u_0,\sigma_{u,j})=\frac{u_j}{\sigma_{u,j}^2}\exp\!\left[-\frac{u_j^2+u_0^2}{2\sigma_{u,j}^2}\right]I_0\!\left(\frac{u_ju_0}{\sigma_{u,j}^2}\right),
 \label{eq:rice}
\end{equation}
where $I_0$ is the modified Bessel function of order zero. This measurement kernel remains correct when the signal-to-noise ratio of $u$ is low.

\paragraph{Good and outlier components.} Not all pairs in the sample are bound binaries of two single stars. With $\texttt{R\_chance\_align}\le0.01$ chance alignments are negligible, but some components are unresolved binaries whose companion is too faint to be detected in the XP spectrum (Sect.~\ref{sec:jcaps}). For these hierarchical triples and quadruples the relation between $u$ and the masses of the two visible stars does not hold: the unseen companion adds mass, and the orbital motion of the inner pair shifts the photocentre measured by Gaia and adds a spurious proper motion. We therefore describe each pair by a mixture of a good component, whose mass scale is that of the two stars, and an outlier component for systems with a hidden inner companion.

\paragraph{Shape of the good component.} The good component uses the functional form of \citet{Hwang2024DynamicalMasses},
\begin{equation}
 p(\ut)=\frac{1}{N}\,\ut\exp\!\left[-B\ut^2-\exp\!\left(\frac{\ut-u_c}{C}\right)\right],
 \label{eq:good}
\end{equation}
with re-calibrated parameters $B=2.6780\times10^{-3}$, $u_c=35.7309$ and $C=3.5811$, and $N$ normalising $p$ on $(0,\infty)$. 

The constants are fitted to a Monte Carlo simulation of Keplerian orbits with uniform mean anomalies and isotropic orientations. The eccentricities follow $p(e)\propto e^{\alpha_e}$ with $\alpha_e=1.23$, the value that \citet{Hwang2022Eccentricity} measured at the separations of our pairs. The semi-major axes follow $\dd N/\dd\ln a\propto a^{-0.6}$, which reproduces the observed distribution of projected separations, $\dd N/\dd s\propto s^{-1.6}$ \citep{ElBadryRix2018}. This distribution matters because the pairs are observed at a given projected separation. When small orbits are more common, a larger fraction of these pairs are eccentric binaries near apocentre, where the relative speed is low. 

Both choices differ from those of \citet{Hwang2024DynamicalMasses}, whose $p(\ut)$ corresponds to $\alpha_e=1$ and a log-uniform distribution of $a$. Under their assumptions our orbit simulation reproduces their distribution. 

The shape is held fixed, and its upper edge is set by energy conservation, $\ut<\sqrt{2G\msun/\mathrm{au}}=42.12\,\uunit$ for any eccentricity distribution. The effect of $\alpha_e$ on the MLR is quantified in Appendix~\ref{sec:sys_pu}.

\paragraph{Shape of the outlier component.} The shape of the outlier component is computed from a forward model of hidden companions (Appendix~\ref{app:companions}). 

For every pair of the sample we give each component an unresolved inner companion with a probability that increases with its mass, and with the period, mass-ratio and eccentricity distributions of field binaries \citep{Raghavan2010,MoeDiStefano2017}. Over the $34$-month baseline of Gaia DR3 the photocentre motion of the inner pair is absorbed into the proper motion when the inner period is comparable to or longer than the baseline, and adds a velocity of up to about $1.5\,\mathrm{km\,s^{-1}}$ to the relative motion of the wide pair, comparable to its orbital velocity. 

Systems in which Gaia would resolve the companion, in which the companion contributes more than $10\%$ of the $G$-band flux and would be detected in the XP spectrum, or which would have a RUWE above $3$, are removed. The retained systems give the distribution of $\ut=u_0/\sqrt{m_{\rm iso}/\msun}$, where $m_{\rm iso}$ is the mass assigned to the two observed, blended magnitudes according to PARSEC MLRs. 

Both this distribution and the fraction of retained systems with a companion depend on the pair. At larger distance a companion is less often resolved and its photocentre motion is less often detected through the RUWE, and the velocity added by the companion enters $u=v\sqrt{s}$ with the factor $\sqrt{s}$ of the wide pair. For each pair $j$ we therefore take the distribution $h_j(\ut)$ and the fraction $r_j$ of retained systems with a companion from the $200$ pairs of the sample that are nearest to it in distance, projected separation and $M_{\mathrm{G}}$ of the primary (Appendix~\ref{app:companions}).

The shapes $h_j$ (Fig.~\ref{fig:template_dist}) are fixed to our simulation. The outlier fraction follows the simulated fractions,
\begin{equation}
 \mathrm{logit}\,f_j=\mathrm{logit}\,f_0+\beta\left(\mathrm{logit}\,r_j-\langle\mathrm{logit}\,r\rangle\right),
 \label{eq:fout}
\end{equation}
where $\langle\cdot\rangle$ is the mean over the sample, and $f_0$ and $\beta$ are free, because the normalisation of the multiplicity statistics and the detection limits of the hidden companions are uncertain. $\beta=1$ corresponds to the variation between pairs predicted by the forward model, and $\beta=0$ to one fraction for all pairs.

As Fig.~\ref{fig:template_dist} shows, $h_j$ extends far beyond the upper limit of $p(\ut)$. Without measurement noise, the density of the scaled velocity of pair $j$ is then
\begin{equation}
 g_j(u_0\mid\kappa_j)=\frac{1}{\kappa_j}\left[(1-f_j)\,p\!\left(\frac{u_0}{\kappa_j}\right)+f_j\,h_j\!\left(\frac{u_0}{\kappa_j}\right)\right],
 \label{eq:mix}
\end{equation}
where the outlier fraction $f_j$ is the probability that pair $j$ has a hidden companion. The outlier term is truncated at $u_0=80$ and renormalised.

The density of the observed $u$ is the convolution of $g_j$ with the measurement kernel,
\begin{equation}
 q_j(u\mid\kappa_j)=\int_0^{\infty}g_j(u_0\mid\kappa_j)\,R(u\mid u_0,\sigma_{u,j})\dd u_0 .
 \label{eq:like}
\end{equation}

\begin{figure*}[t]
\centering
\includegraphics[width=0.87\textwidth]{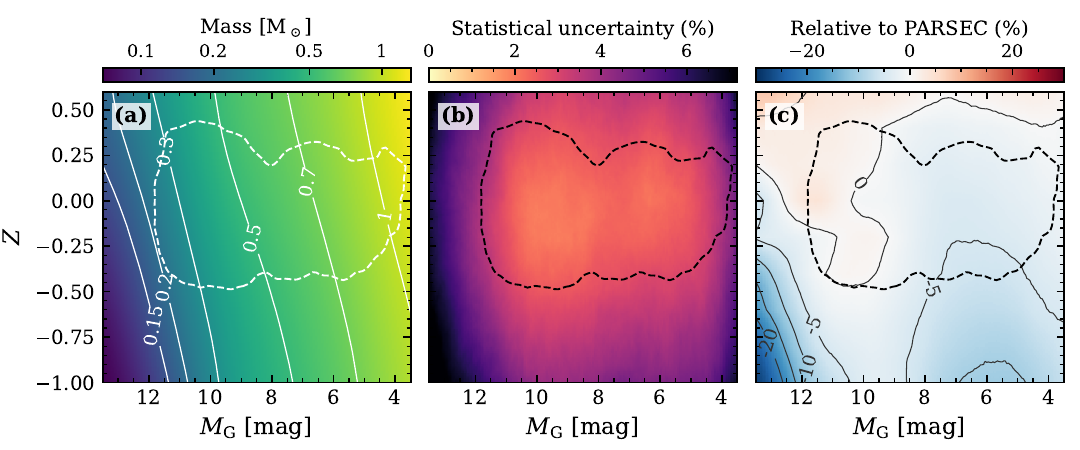}
\caption{Posterior MLR in the plane of $\MG$ and calibrated metallicity $Z$. Panel (a) shows the median mass, and panel (b) its statistical uncertainty. Panel (c) shows the median mass relative to PARSEC at the same $Z$. The dashed contour encloses $95\%$ of the stars of the sample, each placed at its $\MG$ and at the median of $p_j(Z)$ of its pair.}
\label{fig:surface}
\end{figure*}

\subsection{Selection and marginalisation over metallicity}\label{sec:selection}
Our sample retains only pairs whose proper-motion difference is detected at more than $c=3$ times its uncertainty, and the wide-binary catalogue of \citet{ElBadry2021MillionBinaries} includes only pairs whose proper-motion difference is consistent with a bound orbit (Sect.~\ref{sec:data}). Both criteria act on the observed statistic $u_j$ with limits that are fixed for each pair by its astrometric uncertainties, parallax and separation,
\begin{equation}
 c\,\sigma_{u,j}<u_j<U_j,\qquad U_j=65.95+2\sigma^{\rm cat}_{u,j},
 \label{eq:cut}
\end{equation}
where $\sigma^{\rm cat}_{u,j}$ is the uncertainty of Eq.~(\ref{eq:sigu}) computed from the catalogue uncertainties without the inflation, and $U_j$ is converted to the zero-point-corrected distance of Sect.~\ref{sec:data}. For our pairs, $U_j$ lies between $65.7$ and $146\,\uunit$. 

The density of $u_j$ for a pair in the sample is therefore $q_j$ divided by the probability that the pair satisfies both limits,
\begin{equation}
 q^{\rm sel}_j(u_j\mid\kappa_j)=\frac{q_j(u_j\mid\kappa_j)}{\int_{c\,\sigma_{u,j}}^{U_j}q_j(u\mid\kappa_j)\dd u}.
 \label{eq:sel}
\end{equation}
We refer to this division as the selection correction. It is most important at the lower limit, because low-mass pairs have small $u$ and fail the selection more often than massive pairs. A fit without the selection correction would overestimate the masses (Sect.~\ref{sec:tests}).

Finally, the uncertain metallicity of each pair is taken into account by averaging Eq.~(\ref{eq:sel}) over $p_j(Z)$,
\begin{equation}
 \mathcal L_j(\Theta,f_0,\beta)=\int p_j(Z)\,q^{\rm sel}_j\!\left[u_j\mid\kappa_j(Z)\right]\dd Z ,
 \label{eq:marg}
\end{equation}
where $\kappa_j(Z)$ depends on $\Theta$ through Eqs.~(\ref{eq:g}) and (\ref{eq:kappa}). The posterior is the product of the priors and of $\mathcal L_j$ over all pairs, and it has 36 free parameters (Table~\ref{tab:priors}). 

We sample the posterior with the No-U-Turn sampler \citep{HoffmanGelman2014NUTS} as implemented in NumPyro \citep{Phan2019NumPyro}. Four chains converged, with the convergence statistic $\hat R$ below $1.01$ for all parameters \citep{GelmanRubin1992,Vehtari2021}. We refer to this fit as the adopted fit. The priors, the sampler settings and the convergence diagnostics are given in Appendix~\ref{sec:inference}.


\section{Results}\label{sec:results}
\subsection{The MLR and its statistical uncertainty}\label{sec:res_mlr}
The posterior MLR is shown in Fig.~\ref{fig:surface}. The dashed contour encloses $95\%$ of the stars of the sample and extends over $3.7\lesssim\MG\lesssim11.8$\,mag and $-0.5\lesssim Z\lesssim0.4$. We define the statistical uncertainty of the mass as half of the $68\%$ posterior interval. Within the contour this uncertainty lies between $1.9$ and $4.7\%$, with a median of $2.5\%$. Outside the contour the surface is determined mainly by the priors and by the conditions of Eq.~(\ref{eq:mono}).

\begin{figure*}[t]
\centering
\includegraphics[width=0.52\textwidth]{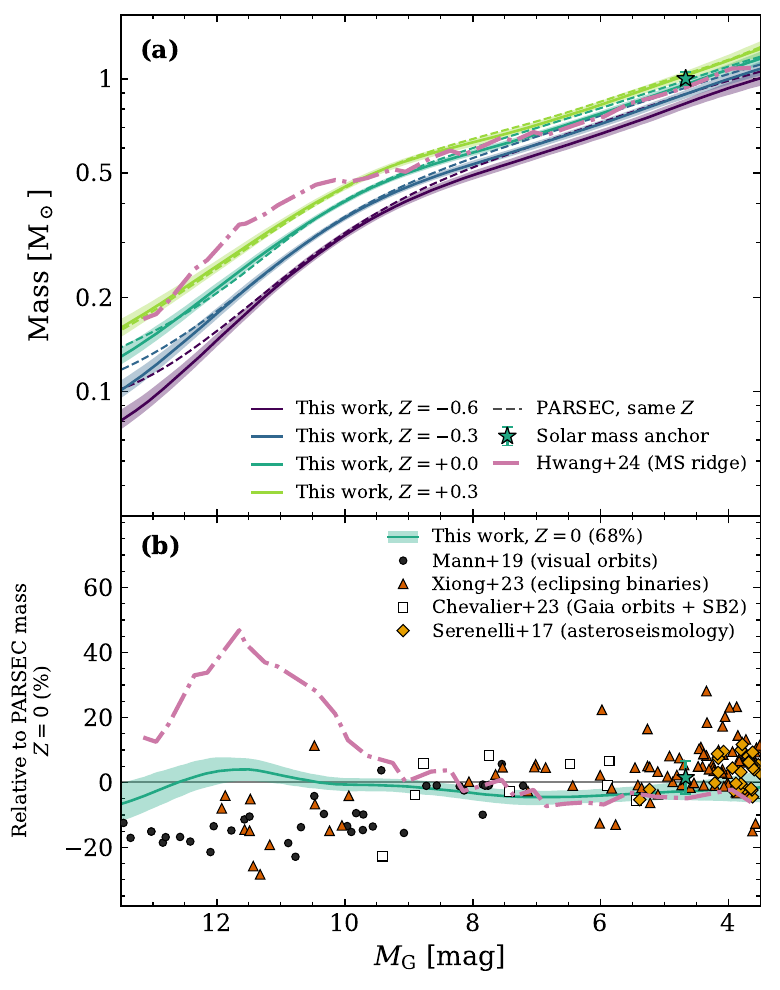}
\caption{The MLR at four metallicities from $Z=-0.6$ to $+0.3$. Panel (a) shows our MLR with its posterior interval (solid lines and bands) and PARSEC at the same metallicities (dashed lines). The dash-dotted line is the relation of \citet{Hwang2024DynamicalMasses} along the ridge of the main sequence, and the star marks the solar mass anchor. Panel (b) shows masses relative to PARSEC at $Z=0$. It includes our MLR at $Z=0$ and the published masses of stars within $0.15$\,dex of solar metallicity (symbols, Sect.~\ref{sec:disc_masses}).}
\label{fig:summary}
\end{figure*}

The MLR is close to that of PARSEC (Fig.~\ref{fig:surface}c), with differences between $-6\%$ and $+4\%$ inside the contour. At solar metallicity the two relations agree to within $5\%$, and our masses are 2 to $5\%$ lower for $\MG=4$ to 8\,mag (Fig.~\ref{fig:summary}). At the solar point we find a mass of $0.957^{+0.027}_{-0.026}\,\msun$. The posterior at this point is about half as wide as the solar mass anchor, so the mass is determined mainly by the dynamics. The masses at $\MG=5$, 8 and 11\,mag are nearly uncorrelated in the posterior (coefficients below $0.15$), so the data determine them largely independently.

\subsection{Metallicity dependence}\label{sec:res_feh}
\begin{figure}[t]
\centering
\includegraphics[width=\columnwidth]{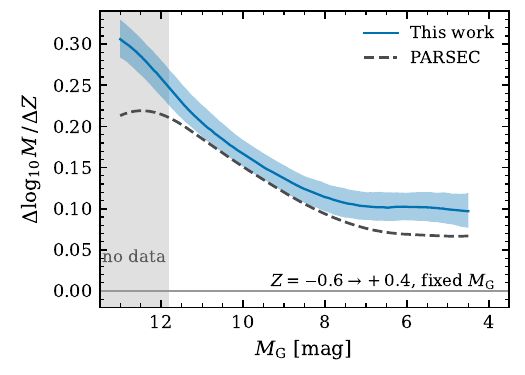}
\caption{Metallicity sensitivity of the mass at fixed $\MG$, measured between $Z=-0.6$ and $+0.4$. The line and band show the posterior median and $68\%$ interval of our MLR, and the dashed line is PARSEC. The data constrain the MLR for $\MG\lesssim11.8$\,mag.}
\label{fig:slope}
\end{figure}

At fixed $\MG$ a metal-rich star is more massive than a metal-poor star, because higher opacity makes it fainter at a given mass \citep[e.g.][]{ChabrierBaraffe2000}. The sign of this dependence is imposed by the model, whereas its size is measured. We express the size by the metallicity sensitivity $\Delta\log_{10}m/\Delta Z$ between $Z=-0.6$ and $+0.4$ (Fig.~\ref{fig:slope}). The sensitivity increases from $0.10\pm0.02$ at $\MG=5$\,mag to $0.20\pm0.02$ at 11\,mag. A change of $0.5$\,dex in metallicity then changes the mass by $12\%$ at $\MG=5$\,mag and by $26\%$ at 11\,mag. Since these changes are several times the statistical uncertainty of the MLR, masses derived from $\MG$ require a metallicity.

The sensitivity of PARSEC increases towards faint magnitudes in the same way. It is lower than ours by $0.02$ to $0.03$, which is 1 to 2 times our statistical uncertainty. In a fit restricted to the pairs with $u/\sigma_u>10$ (Sect.~\ref{sec:tests}), the difference is $0.04$ to $0.05$, or $2\sigma$ to $3\sigma$.

\subsection{Goodness of fit and outlier fraction}\label{sec:res_fit}
\begin{figure}[t]
\centering
\includegraphics[width=0.85\columnwidth]{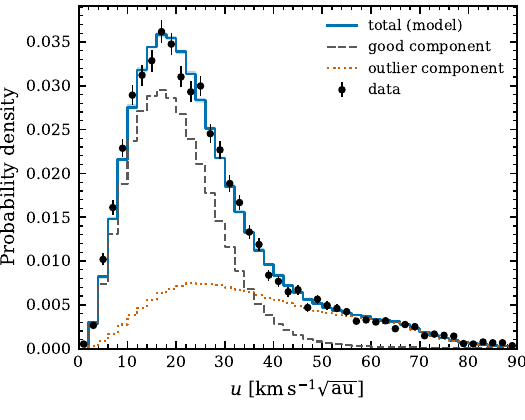}
\caption{Observed distribution of $u$ for the $10\,559$ pairs (points with Poisson errors) and the posterior prediction of the model with the selection (blue line). The dashed and dotted lines are the good component and the outlier component.}
\label{fig:upred}
\end{figure}

Figure~\ref{fig:upred} compares the observed distribution of $u$ with the posterior prediction of the model after the selection. The model reproduces both the core of the distribution, where the good component dominates, and the extended tail, which is due to the outlier component. This tail falls off between $u=66$ and 80, where the upper limit of the selection acts and the outlier term is truncated. At small $u$ the outlier component also contributes, and there it overlaps with the good component.

Before the selection, the fitted outlier fraction $f_j$ has a mean of $0.29$. The outlier component lies at larger $u$ than the good component, so its pairs satisfy the lower limit more often. In the posterior prediction they make up one third of the sample. The fraction $f_j$ increases with distance, by a factor of two between the nearest and the most distant third of the pairs. This increase is steeper than predicted by the companion simulation ($\beta=1.37\pm0.14$, compared with 1). In the posterior, a larger mean outlier fraction corresponds to slightly lower masses at $\MG=5$ and 8\,mag (correlation coefficients of $-0.26$ and $-0.35$).

The steeper increase may indicate that hidden companions of distant pairs escape detection more easily than we assume. Their XP spectra are noisier, and a given photocentre motion corresponds to a smaller angle on the sky. Alternatively, distant pairs may contain outliers of a different origin. In either case the simulation predicts fewer outliers than the fit requires (Appendix~\ref{app:companions}). The simulation is thus a better description of the shape of the outlier tail than of its amplitude.

\begin{figure*}[t]
\centering
\includegraphics[width=0.7\textwidth]{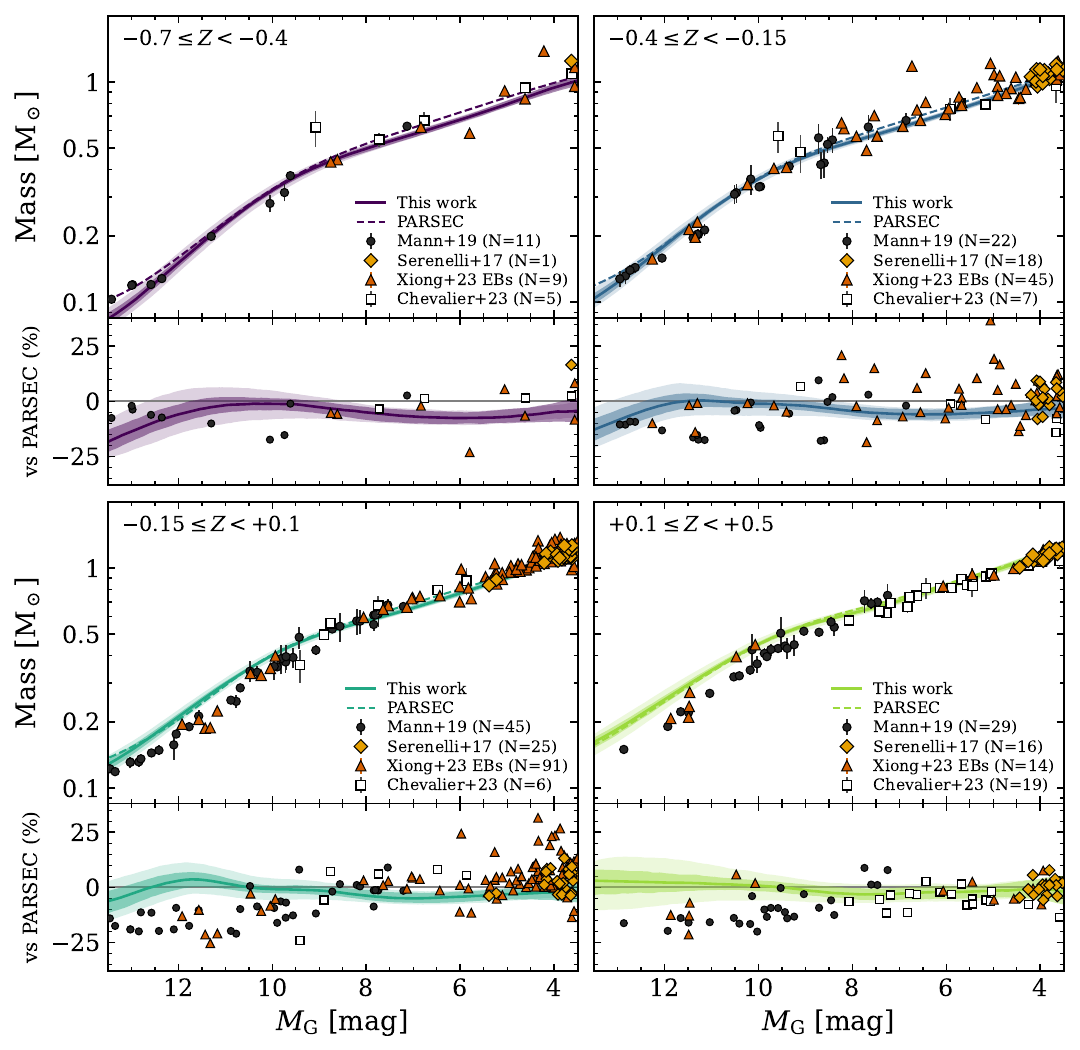}
\caption{MLR in four bins of the calibrated metallicity $Z$. In each bin, the upper panel shows the mass and the lower panel the difference from PARSEC in per cent. The solid line is our MLR at the bin centre with its $68\%$ interval (dark band); the light band spans our MLR between the two bin edges. The dashed line is PARSEC at the bin centre. The published samples of Fig.~\ref{fig:summary} are binned by their own metallicity ($[\mathrm{Fe/H}]$ for \citealt{Mann2019} and \citealt{Xiong2023}, $[\mathrm{M/H}]$ for \citealt{Serenelli2017}), and the Gaia orbits of \citet{Chevalier2023} (open squares) by their calibrated J-CaPS metallicity. The legends give the number of stars in each bin. In the lower panels each star is compared with PARSEC at its own metallicity.}
\label{fig:mlr_feh_bins}
\end{figure*}

\section{Discussion}\label{sec:discussion}
\subsection{Comparison with individual mass measurements}\label{sec:disc_masses}
We compare the MLR with four published samples of stars whose masses were measured by other methods (Fig.~\ref{fig:summary}). The samples are visual orbits of M dwarfs \citep{Mann2019}, eclipsing binaries \citep{Xiong2023}, Gaia astrometric and spectroscopic orbits \citep{Chevalier2023}, and asteroseismic dwarfs \citep{Serenelli2017}. Gaia measures $\MG$ directly for the Gaia orbits and the asteroseismic dwarfs. The stars of these two samples are all brighter than $\MG=10.5$\,mag. For the visual orbits and eclipsing binaries, we convert the published $K_s$ magnitudes or luminosities to $\MG$ with PARSEC, so their $\MG$ is model dependent.

The metallicity scales of these samples differ from ours, so a comparison at a given metallicity is approximate. We select the stars within $0.15$\,dex of solar metallicity on the scale of each sample and compare them with our MLR at $Z=0$. The Gaia orbits have no published metallicity, and for them we use calibrated J-CaPS metallicities.

The published masses decrease more steeply with $\MG$ than our MLR. For $\MG<8$\,mag their median is 3 to $11\%$ above our MLR, whereas between 8 and $10.5$\,mag it is 2 to $9\%$ below. For fainter stars the median is about $17\%$ below our MLR. The relation of \citet{Hwang2024DynamicalMasses} lies far above our MLR and the published masses at the faint end, probably because unresolved binaries remain in their sample.

\begin{figure*}[t]
\centering
\includegraphics[width=0.92\textwidth]{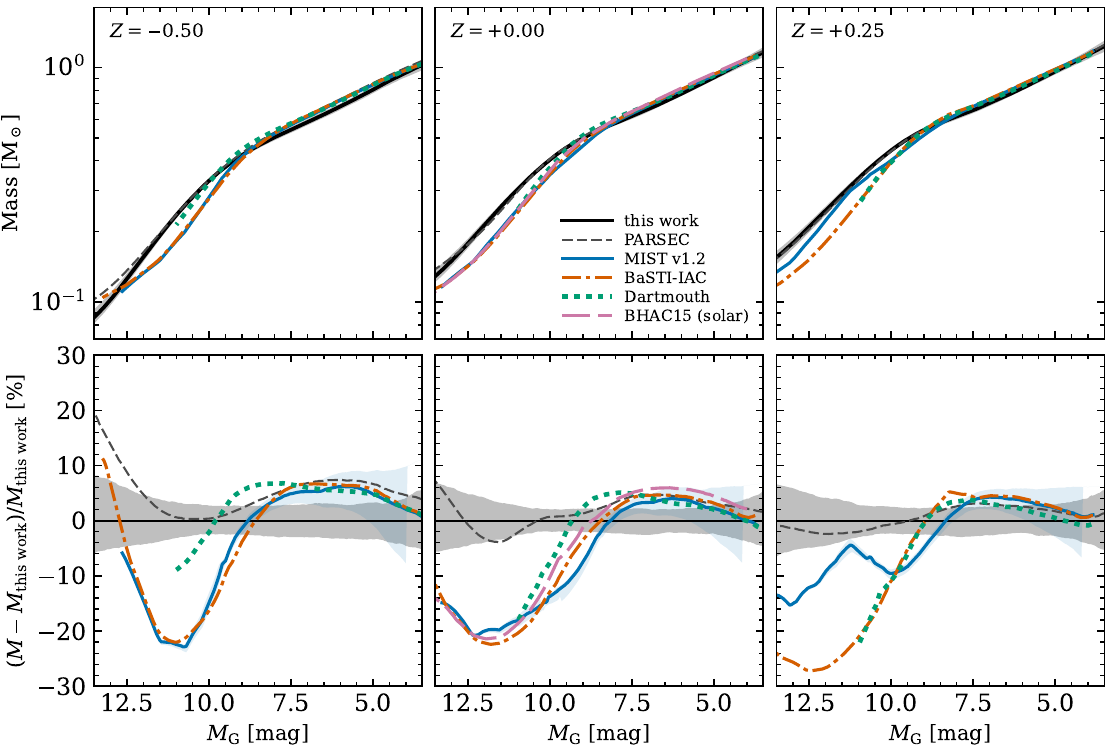}
\caption{MLR and stellar models at $Z=-0.5$, 0 and $+0.25$. The top row shows the mass. The bottom row shows the difference from the median of our MLR in per cent, and the grey band is our $68\%$ interval. Coloured lines are isochrones of 5\,Gyr, and the dashed line is PARSEC at 4\,Gyr, the age of our prior. The light-blue band is the range of MIST for ages of 2 to 10\,Gyr. BHAC15 is available for solar composition only, and the Dartmouth curve is cut at $\MG=11$\,mag, where its mass grid becomes too coarse.}
\label{fig:models}
\end{figure*}

Figure~\ref{fig:mlr_feh_bins} repeats the comparison in four bins of $Z$. Here the published samples are binned by their own metallicity, and the Gaia orbits by their calibrated J-CaPS metallicity, so that all stars are used and not only those near solar metallicity. At $\MG<8$\,mag the published masses scatter around our MLR in all bins. At fainter magnitudes, the visual orbits of the most metal-poor bin agree with our MLR to within their scatter. In the three more metal-rich bins, the visual orbits and eclipsing binaries at $\MG\gtrsim9$\,mag lie about 10 to $20\%$ below PARSEC, whereas our MLR stays within about $5\%$ of PARSEC. The faint-end difference therefore appears at all metallicities above $Z\approx-0.4$. Because the metallicity scales of the samples differ from ours, the assignment of individual stars to the bins is approximate.

The difference at the faint end is the main open question of this paper. One possible cause is the conversion to $\MG$. Only the visual orbits and the eclipsing binaries reach these magnitudes, and both rely on PARSEC for it. Near $\MG=11$\,mag the mass changes by about $3\%$ per $0.1$\,mag. The converted magnitudes would have to be too bright by about $0.5$\,mag to remove the difference.

The $K_s$ conversion is sensitive to metallicity, by about $0.9$\,mag per dex. We tested it for the sample of \citet{Mann2019} with the Gaia $G$ magnitudes, although these are not reliable for close, blended pairs. The resulting $\MG$ differs from the converted value by about $0.14$\,mag on average, or about $4\%$ in mass. With these magnitudes the measured total masses of the faint binaries remain 12 to $16\%$ below our prediction. For this sample the $K_s$ conversion thus explains at most a third of the difference.

Alternatively, our MLR may be too high at the faint end. The published masses at $\MG>10.5$\,mag are 15 to $16\%$ below PARSEC. MIST and BaSTI are 13 to $20\%$ below PARSEC at these magnitudes (Sect.~\ref{sec:disc_models}), close to the published masses. Our faint-end masses are also sensitive to the outlier model (Sect.~\ref{sec:tests}), so residual errors of the outlier distributions could bias them. The agreement of our MLR with PARSEC there is not imposed by the prior on $D$, which would permit a difference of this size.

\subsection{Comparison with stellar models}\label{sec:disc_models}
Figure~\ref{fig:models} compares our MLR with four further sets of stellar models, namely MIST \citep{Dotter2016MIST,Choi2016MIST}, BaSTI-IAC \citep[][BaSTI hereafter]{Hidalgo2018BaSTI}, Dartmouth \citep{Dotter2008Dartmouth} and BHAC15 \citep{Baraffe2015}. We use 5\,Gyr isochrones with the Gaia magnitudes published by each group, and we identify the model $[\mathrm{Fe/H}]$ with $Z$. At solar metallicity and $4\lesssim\MG\lesssim8$\,mag, all five sets of models agree with each other and with our MLR to within about $6\%$. Our masses are a few per cent below the models in this range. The models diverge at fainter magnitudes.

For $\MG\ge10$\,mag, MIST and BaSTI are 13 to $20\%$ below PARSEC and 13 to $22\%$ below our MLR. BHAC15 follows the same trend, and Dartmouth is $8\%$ below our MLR at $\MG=10$\,mag. PARSEC, in contrast, remains within $5\%$ of our MLR at solar metallicity over the range with data. At the other two metallicities these models also lie below our MLR at faint magnitudes (Fig.~\ref{fig:models}).

Age is unlikely to explain the divergence at faint magnitudes. For the MIST isochrones, a change of the age from 5\,Gyr to 2 or 10\,Gyr changes the mass by at most $3\%$ for $\MG\ge6$\,mag. At the bright end the effect of age is larger. It reaches 4 to $5\%$ at $\MG=5$\,mag, which is as large as our offset from the models there.

The divergence instead arises from differences in the equation of state, the opacities, the treatment of convection and the atmospheric boundary condition. PARSEC, in particular, applies an empirical correction to this boundary condition that was calibrated on the observed mass--radius relation of low-mass stars \citep{Chen2014}. The synthetic photometry that converts luminosity to $\MG$ also differs between the sets. Part of the differences in Fig.~\ref{fig:models} may therefore be photometric, and we cannot separate this part from the interior models.

The metallicity scale affects any comparison of this kind. For a sensitivity of $0.1$ to $0.2$, a shift of $0.1$\,dex between two scales changes the mass by 2 to $5\%$. This is as large as the differences between our MLR and the models for $\MG\lesssim8$\,mag. A shift of this size can arise from our zero point or from the spectroscopic scales of the published samples. Solar-scaled models can also cause it, because they underestimate the total metallicity of $\alpha$-enhanced stars by up to about $0.2$\,dex. MLRs can therefore be compared at the level of a few per cent only if they are on a common metallicity scale.

\subsection{Sensitivity to model assumptions}\label{sec:tests}
\begin{table}
\caption{Change of the median mass at $Z=0$ in per cent when one element of the fit is changed.}
\label{tab:tests}
\centering
\small
\setlength{\tabcolsep}{4.5pt}
\begin{tabular}{@{}ccccccc@{}}
\toprule
$\MG$ & Stat. & $u/\sigma_u$ & $d<250$ & One & No upper & No lower \\
(mag) & unc. & $>10$ & pc & shape & limit & limit \\
\midrule
4  & 3.7 & $+1.3$ & $+0.5$ & $+7.6$ & $+1.9$ & $-1.5$ \\
6  & 2.3 & $+4.6$ & $+2.1$ & $+2.4$ & $+2.2$ & $+12.0$ \\
8  & 2.2 & $+3.4$ & $+1.7$ & $-5.8$ & $+2.0$ & $+27.2$ \\
10 & 1.9 & $-0.1$ & $-0.7$ & $-12.2$ & $+1.2$ & $+29.4$ \\
12 & 3.7 & $-1.1$ & $-0.9$ & $-12.0$ & $+0.4$ & $+18.4$ \\
\bottomrule
\end{tabular}
\tablefoot{A positive value is a higher mass than in the adopted fit. The second column is the statistical uncertainty of the adopted fit. The third and fourth columns are fits of subsamples. ``One shape'' uses one outlier distribution and one outlier fraction for all pairs. The last two columns omit the selection correction for that limit. The largest changes quoted in the text can lie between the rows.}
\end{table}

To test the model assumptions, we repeated the fit with one element changed at a time (Table~\ref{tab:tests}; details in Appendix~\ref{sec:systematics}). Pairs with low $u/\sigma_u$ are the main source of systematic uncertainty. A fit restricted to the $3\,862$ pairs with $u/\sigma_u>10$ gives masses up to $5\%$ higher for $\MG\le8$\,mag and nearly unchanged masses for fainter stars. A fit of the pairs within 250\,pc shows that distant pairs cause about half of this change.

Distant pairs with low $u/\sigma_u$ have the largest $\sigma_u$ and outlier fractions, so the measurement kernel and the outlier model are most important for them. We therefore take the change in the fit with $u/\sigma_u>10$ as the systematic uncertainty of these two parts of the model. In this restricted fit the masses for $\MG=4$ to 8\,mag are 0 to $3\%$ below PARSEC, compared with 2 to $5\%$ in the adopted fit. The restricted fit also has a higher metallicity sensitivity (Sect.~\ref{sec:res_feh}).

The selection correction is the largest single effect. If it is omitted for the lower limit, the masses increase by up to $31\%$, with the largest change at $\MG=8$ to 10\,mag. The MLR therefore depends on an accurate measurement kernel near the lower limit, and in particular on the correct size of $\sigma_u$. The posterior prediction with the selection reproduces the observed distribution of $u$ (Fig.~\ref{fig:upred}), which is consistent with the adopted $\sigma_u$. The upper limit has a much smaller effect of about $2\%$ or less, which is nearly the same at all metallicities.

The outlier model must depend on the distance and separation of the pairs. If a single outlier distribution and a single outlier fraction are used for all pairs, the masses change by up to $12\%$. Such a fit fails to reproduce the $u$ distributions of nearby and distant pairs separately. Its cumulative distributions differ from the data by almost twice the sampling noise, whereas those of the adopted fit stay within it. For this reason we do not treat this change as an uncertainty. It nevertheless shows that the masses are sensitive to the outlier model, in particular at the faint end.

The other tests have small effects. A different eccentricity index, $\alpha_e=1.0$ or $1.5$ in place of $1.23$, changes the masses by at most $2\%$. Without the solar mass anchor, the masses at the bright end are lower by less than their statistical uncertainty (Fig.~\ref{fig:anchor}). For $\MG\gtrsim8$\,mag this fit leaves the MLR unchanged. None of these tests changes the metallicity sensitivity.

\subsection{Caveats}\label{sec:caveats}
The MLR is most sensitive to the outlier model and to the selection correction, in particular for pairs with low $u/\sigma_u$ (Sect.~\ref{sec:tests}). About a third of the sample belongs to the outlier component, whose distributions are taken from a simplified simulation. The selection correction is large and depends on the adopted $\sigma_u$.

The metallicity distributions of the pairs are held fixed in the dynamical fit, so the uncertainty of the metallicity calibration is not propagated. In addition, the zero point of the $Z$ scale is an assumption, and the absolute magnitudes are treated as exact. Because age is not a variable of the model, the MLR is the mean relation for the mixture of ages in the sample. Finally, $p(\ut)$ depends on the adopted distributions of eccentricity and semi-major axis (Sect.~\ref{sec:dynamics}).

\subsection{Prospects with Gaia DR4}\label{sec:dr4}
The outlier distribution and the mass scale are partly degenerate, because a heavier outlier tail can be compensated by a lower mass scale. We fixed the outlier distribution with a simulation for this reason. Dynamical masses from orbits would remove the degeneracy if they were included in the likelihood. The orbital solutions of Gaia DR4 should cover many more binaries and a wider range of $\MG$ and metallicity than the sample of \citet{Chevalier2023}. At faint magnitudes they would test our MLR with $\MG$ measured directly by Gaia, and the epoch astrometry should identify many of the hidden companions.

\section{Conclusions}\label{sec:conclusions}
We have calibrated the MLR of single main-sequence stars in the Gaia $G$ band as a function of metallicity. The calibration is based on the relative speeds and separations of $10\,559$ wide binaries, which constrain the total mass of a pair statistically. We calibrated the metallicities from Gaia XP spectra with the pairs themselves, because the two stars of a pair have the same metallicity. In our model the MLR is a smooth, monotone surface in $\MG$ and metallicity. The likelihood includes the measurement noise, both selection limits, and pairs with a hidden companion, whose effect we simulate. Our main results are as follows.
\begin{enumerate}
 \item The data constrain the MLR over $3.7\lesssim\MG\lesssim11.8$\,mag and $-0.5\lesssim Z\lesssim0.4$. In this region the statistical uncertainty of the mass is between $1.9$ and $4.7\%$, with a median of $2.5\%$. The mass scale is set by the dynamics, and a weak prior at the solar point has little effect on it.
 \item At solar metallicity the MLR agrees with PARSEC to within $5\%$. For $\MG=4$ to 8\,mag all five sets of stellar models agree with the MLR to within about $6\%$. The models diverge for fainter stars, where MIST and BaSTI predict masses 13 to $22\%$ lower than the MLR for $\MG\ge10$\,mag.
 \item At fixed $\MG$ the mass increases with metallicity. The sensitivity $\Delta\log_{10}m/\Delta Z$ increases from $0.10$ at $\MG=5$\,mag to $0.20$ at 11\,mag. A change of $0.5$\,dex in metallicity changes the mass by 12 to $26\%$. This dependence is $1\sigma$ to $2\sigma$ stronger than PARSEC predicts, and it requires MLRs to be compared on a common metallicity scale.
 \item Published masses of individual stars are 3 to $11\%$ above our MLR for $\MG<8$\,mag and about $17\%$ below it for $\MG>10.5$\,mag. For the visual orbits, the conversion of the published magnitudes to the $G$ band explains at most a third of the faint-end difference. The origin of this difference is not yet understood, and our MLR may be too high there.
 \item About a third of the sample belongs to the outlier component. It consists mainly of pairs in which a hidden companion, undetected in the XP spectrum and through the RUWE, moves the photocentre of one star. A simulation of such companions reproduces the shape of this component but predicts a lower amplitude than the fit requires.
 \item The selection correction for the lower limit is essential, because without it the masses would be up to $31\%$ higher. It requires an accurate model of the measurement noise. The largest systematic uncertainty comes from the pairs with low $u/\sigma_u$, since a fit restricted to $u/\sigma_u>10$ gives masses up to $5\%$ higher.
\end{enumerate}
Gaia DR4 is expected to provide dynamical masses for many more binaries, with magnitudes measured in the $G$ band. These masses will test the faint end of the MLR and can be included in the likelihood as independent constraints on the mass scale.

\section*{Data availability}\label{sec:release}
The MLR is available at \url{https://github.com/yt-wangg/wide-binary-mlr} as the Python package \texttt{wbmlr}. The package provides the median mass and its $68\%$ interval as functions of $\MG$ and $Z$. It also flags the region constrained by the data and provides $4\,000$ posterior samples of the spline coefficients. Two tables in the repository give the MLR on a grid of $\MG$ and $Z$ and the mass on the colour--magnitude diagram. For the second table, the MLR is averaged over the metallicities of the sample stars at each position. The code and the input data of the fit are available at \url{https://github.com/yt-wangg/wide-binary-mlr-fit}.

\begin{acknowledgements}
This work has made use of data from the European Space Agency (ESA) mission Gaia (\url{https://www.cosmos.esa.int/gaia}), processed by the Gaia Data Processing and Analysis Consortium (DPAC, \url{https://www.cosmos.esa.int/web/gaia/dpac/consortium}). Funding for the DPAC has been provided by national institutions, in particular the institutions participating in the Gaia Multilateral Agreement. Y.-S.T. acknowledges support from a fellowship of the Alexander von Humboldt Foundation. We used the software packages NumPyro \citep{Phan2019NumPyro}, JAX \citep{Bradbury2018JAX}, Astropy \citep{Astropy2013,Astropy2018,Astropy2022}, Matplotlib \citep{Hunter2007}, NumPy \citep{Harris2020NumPy} and SciPy \citep{Virtanen2020SciPy}. The authors used Claude Code with Claude Sonnet~5.5, Claude Opus~5.5 and Claude Fable~5.2 (Anthropic) and Codex with GPT-6 Sol (OpenAI) for code development, drafting and editing of the text. The authors reviewed the results and the text and take full responsibility for the content.
\end{acknowledgements}

\bibliographystyle{aa}
\bibliography{refs}

\begin{appendix}
\nolinenumbers

\makeatletter
\twocolumn[{%
\setcounter{section}{1}%
\begin{minipage}{\textwidth}\def\@captype{table}%
\centering
\caption{Parameters and priors. Normal distributions are given as location and standard deviation.}
\label{tab:priors}
\small
\begin{tabular}{@{}lll@{}}
\toprule
Quantity & Prior & Role \\
\midrule
$b$, $s_z$, $\nu$ & maximum likelihood & metallicity calibration, Eq.~(\ref{eq:zdiff}); held fixed \\
$b$ at $\MG=4.67$\,mag & fixed to $0$ (assumed) & zero point of the metallicity scale \\
$\bm\omega$ & Dirichlet$(1,\ldots,1)$ & mixture fractions of $p_{\rm pop}$ \\
$c_0$ & $\Normal(\Pi_{11},0.10)$\,dex & first coefficient \\
$a_l,b_i,e_{il}$ & $\Normal(0,1)$ & variables of the monotone grid \\
$\ln\lambda_M$ & $\Normal(\ln\lambda_{M,\rm ref},0.7)$ & increment scale in $M_{\mathrm{G}}$ \\
$\ln\lambda_Z$ & $\Normal(\ln\lambda_{Z,\rm ref},0.7)$ & increment scale in $Z$ \\
$\mathrm{logit}\,f_0$ & $\Normal(\mathrm{logit}\,0.2,1)$ & outlier fraction at the mean of $\mathrm{logit}\,r_j$ \\
$\beta$ & $\Normal(1,0.5)$ & variation of the outlier fraction with $r_j$ \\
$D_{il}=\Theta_{il}-\Pi_{il}$ & $\Normal(0,\tau_D=0.10)$\,dex & shrinkage to PARSEC \\
differences of $D$, Eq.~(\ref{eq:smooth}) & $\Normal(0,\tau=0.05)$\,dex & smoothness in $M_{\mathrm{G}}$, $Z$ and mixed \\
$\log_{10}m$ at the solar point & $\Normal(0,0.05/\ln10)$\,dex & solar mass anchor ($1\,\msun\pm5\%$) \\
\bottomrule
\end{tabular}
\end{minipage}\par\vspace{1.5\baselineskip}
\begin{minipage}{\textwidth}\def\@captype{figure}%
\centering
\includegraphics[width=0.95\textwidth]{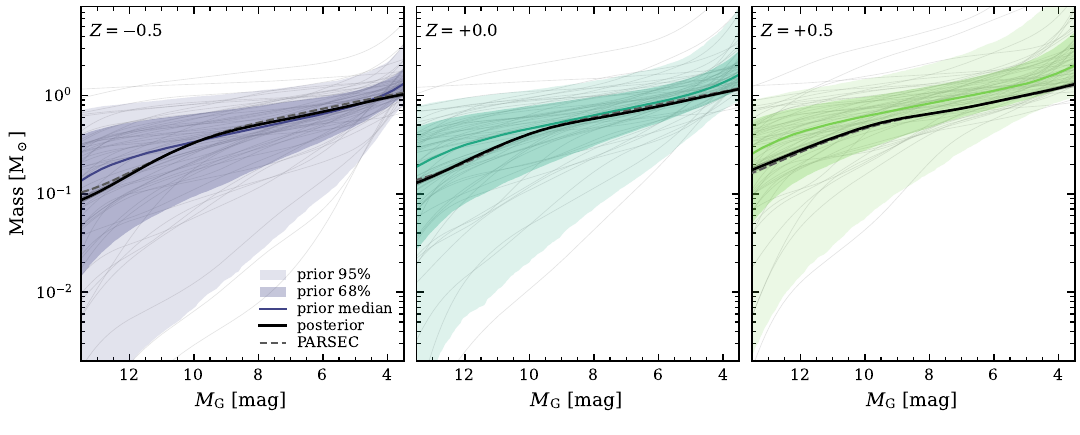}
\caption{Prior of the MLR surface projected onto the mass--$M_{\mathrm{G}}$ plane at three metallicities. The shaded bands show the $68\%$ and $95\%$ ranges of $4\,000$ draws from the priors on the sampled coordinates of Table~\ref{tab:priors}, the coloured line is their median, and a subset of the draws is drawn in grey. The shrinkage, smoothness and solar mass anchor terms act on the derived surface and are not included. The black line and band are the median and $68\%$ interval of the posterior, and the dashed line is PARSEC at the same metallicity. Every prior draw is monotone in $M_{\mathrm{G}}$ and $Z$ by construction.}
\label{fig:prior_pred}
\end{minipage}\par\vspace{\baselineskip}
\setcounter{section}{0}%
}]
\makeatother

\section{Model specification}\label{app:model}
\paragraph{Spline surface.} Equation~(\ref{eq:g}) describes $\log_{10}m$ by $8\times4$ coefficients $\Theta_{il}$, one for each product of a B-spline $\phi_i$ in $\MG$ and a B-spline $\psi_l$ in $Z$. Both sets of B-splines are cubic. In $\MG$ their knots lie at $3.5$, $5.5$, $7.5$, $9.5$, $11.5$ and $13.5$\,mag, with the outer knots repeated, which gives eight B-splines. In $Z$ there are no interior knots, so the four B-splines are cubic polynomials on $-1.0\le Z\le0.6$, and at fixed $\MG$ the surface is a cubic function of $Z$.

Each B-spline is non-zero only over a limited range and is largest near its centre, the mean of the three inner knots of its support. These centres lie at $\MG=3.5$, $4.2$, $5.5$, $7.5$, $9.5$, $11.5$, $12.8$ and $13.5$\,mag and at $Z=-1.0$, $-0.47$, $+0.07$ and $+0.6$. The coefficient $\Theta_{il}$ is approximately the value of $\log_{10}m$ at the centres of $\phi_i$ and $\psi_l$. The coefficients thus form a coarse grid of $\log_{10}m$ in the plane of $\MG$ and $Z$, with $i$ increasing towards fainter stars and $l$ towards higher metallicity, and the surface is a smooth interpolation of this grid.

\paragraph{Monotone parametrisation.} The conditions of Eq.~(\ref{eq:mono}) state that the coefficients decrease along every row of the grid towards fainter stars and increase along every column towards higher metallicity. The derivative of a B-spline surface along $\MG$ is a sum of non-negative basis functions weighted by the differences of neighbouring coefficients along $\MG$, and likewise for $Z$. If the coefficients satisfy Eq.~(\ref{eq:mono}), the mass therefore decreases with $\MG$ and increases with $Z$ everywhere in the domain, not only at the grid points.

To fit the MLR we do not sample the coefficients directly, because most combinations would violate these conditions. Instead we build the grid from variables that can take any real value, in such a way that every grid satisfies Eq.~(\ref{eq:mono}). Two functions turn these variables into allowed steps: $\softplus(t)=\ln(1+e^t)$, which is always positive, and $\sigmoid(t)=(1+e^{-t})^{-1}$, which lies between 0 and 1. We start at the corner of the grid for the brightest and most metal-poor stars and fill the two edges that meet there,
\begin{align}
 \Theta_{11}&=c_0,\\
 \Theta_{1l}&=\Theta_{1,l-1}+\lambda_Z\softplus(a_l),\quad l=2,\ldots,4,\\
 \Theta_{i1}&=\Theta_{i-1,1}-\lambda_M\softplus(b_i),\quad i=2,\ldots,8.
\end{align}
Along the edge at $\MG=3.5$\,mag every step to higher metallicity adds a positive amount, and along the edge at $Z=-1.0$ every step to fainter stars subtracts one. The positive scales $\lambda_M$ and $\lambda_Z$ set the typical size of a step. Each interior coefficient must not exceed its brighter neighbour $\Theta_{i-1,l}$ and must not fall below its more metal-poor neighbour $\Theta_{i,l-1}$. We therefore place it between the two,
\begin{multline}
 \Theta_{il}=\Theta_{i,l-1}+\left(\Theta_{i-1,l}-\Theta_{i,l-1}\right)\sigmoid(e_{il}),\\
 i=2,\ldots,8,\quad l=2,\ldots,4,
\end{multline}
and fill the grid in order of increasing $i+l$, so that both neighbours are known when a coefficient is set. The interval is never empty, because the two neighbours already satisfy the conditions and hence $\Theta_{i,l-1}\le\Theta_{i-1,l-1}\le\Theta_{i-1,l}$. The variables $c_0$, $a_l$, $b_i$ and $e_{il}$ give $1+3+7+21=32$ coordinates, and together with $\ln\lambda_M$ and $\ln\lambda_Z$ they are the sampled parameters of the MLR. Every posterior draw satisfies Eq.~(\ref{eq:mono}), and no draw has to be rejected.

\paragraph{Priors.} The priors of the MLR have two parts. The first part acts on the sampled variables and is chosen so that a typical draw resembles PARSEC. The second part acts on the coefficients $\Theta_{il}$ of the surface and keeps the fitted relation close to PARSEC and smooth where the data are few. Table~\ref{tab:priors} lists all parameters and priors, including those of the metallicity calibration.

The corner coefficient $c_0$ has a Gaussian prior centred on the PARSEC value $\Pi_{11}=0.00703$ with a width of $0.10$\,dex. The variables $a_l$, $b_i$ and $e_{il}$ have standard normal priors. For $a_l=b_i=0$ a step along an edge equals $\lambda_Z\ln2$ or $\lambda_M\ln2$, and for $e_{il}=0$ an interior coefficient lies midway between its two neighbours.

The scales $\lambda_M$ and $\lambda_Z$ have log-normal priors with a width of $0.7$, centred on $\lambda_{M,\rm ref}=0.18624$ and $\lambda_{Z,\rm ref}=0.13078$\,dex. These values make a step with $a_l=0$ or $b_i=0$ equal to the median step between neighbouring PARSEC coefficients along $\MG$ and along $Z$. Together with $f_0$ and $\beta$ of the outlier fraction (Eq.~(\ref{eq:fout})), the MLR fit thus has $36$ free parameters.

The second part consists of three terms. Each is added to the log-posterior as a function of $\Theta$ and is not a prior on a single sampled variable. We write $\Theta=\Pi+D$ as in Sect.~\ref{sec:mlr}. The shrinkage term,
\begin{equation}
 \log\pi_{\rm shrink}=-\sum_{il}\frac{D_{il}^2}{2\tau_D^2},\qquad\tau_D=0.10\,{\rm dex},
\end{equation}
pulls every coefficient of the correction towards zero, that is, the surface towards PARSEC. The smoothness term penalises corrections that change abruptly from one grid point to the next \citep{EilersMarx1996PSplines},
\begin{multline}
 \log\pi_{\rm smooth}=-\sum_{il}\frac{(\Delta^2_{M}D_{il})^2}{2\tau_M^2}-\sum_{il}\frac{(\Delta^2_{Z}D_{il})^2}{2\tau_Z^2}\\-\sum_{il}\frac{(\Delta_{MZ}D_{il})^2}{2\tau_{MZ}^2},
 \label{eq:smooth}
\end{multline}
with $\tau_M=\tau_Z=\tau_{MZ}=0.05$\,dex. Each sum runs over the coefficients for which the difference is defined. On an evenly spaced grid, $\Delta^2_{M}D_{il}=D_{i+1,l}-2D_{il}+D_{i-1,l}$ is the second difference along $\MG$, and it measures how much the slope of the correction changes. $\Delta^2_{Z}D_{il}$ is the same along $Z$, and the mixed difference $\Delta_{MZ}D_{il}=D_{i+1,l+1}-D_{i+1,l}-D_{i,l+1}+D_{il}$ measures how much the metallicity dependence of the correction changes with $\MG$.

The centres of the B-splines are not evenly spaced near the edges of the grid. We therefore compute each difference as the corresponding derivative at the centres and multiply it by the square of the mean spacing of the centres along that axis, or, for the mixed difference, by the product of the mean spacings along $\MG$ and $Z$. For evenly spaced centres this gives the differences above.

The third term is the solar mass anchor of Sect.~\ref{sec:mlr}, a Gaussian prior with mean $0$ and width $0.05/\ln10\approx0.022$\,dex on $\log_{10}m$ at the absolute magnitude and metallicity of the Sun. None of the three terms constrains the sign of a slope or of a curvature. The signs of the slopes follow from the monotone parametrisation alone.

The priors are weak where the data are dense. Figure~\ref{fig:prior_pred} shows the surfaces drawn from the priors on the sampled variables. At faint magnitudes they span more than an order of magnitude in mass, far more than the posterior. The solar mass anchor is the tightest term, but it acts at a single point. Inside the region covered by the data the MLR is therefore determined mainly by the data. Outside it, the shrinkage and smoothness terms and the monotonicity determine it.

\paragraph{Metallicity prior.} The metallicity prior $p_{\rm pop}(Z)$ of Eq.~(\ref{eq:zpost}) is a sum of six Gaussians with fixed, equally spaced centres from $-1.0$ to $0.6$\,dex. Each Gaussian has a standard deviation of $0.27$\,dex and is truncated to this range. Because the Gaussians overlap, their sum is smooth, and its terms have no physical meaning. We did not tune their number. We give the six fractions $\bm\omega$ a flat prior, sample them in a separate fit before the dynamical fit, and average $p_j(Z)$ over their posterior.

\section{Simulation of hidden companions}\label{app:companions}
The simulation provides the outlier distributions $h_j$ and the fractions $r_j$ of Sect.~\ref{sec:dynamics}. It places hidden companions in many realisations of each observed pair and computes the scaled velocity that would be measured. In each realisation a star receives a companion with a probability that increases with its mass. The periods, mass ratios and eccentricities of the companions follow the statistics of field binaries \citep{Raghavan2010,MoeDiStefano2017}. Companions whose orbit would be unstable within the wide pair are omitted \citep{MardlingAarseth2001}.

\begin{figure}[!tb]
\centering
\includegraphics[width=\columnwidth]{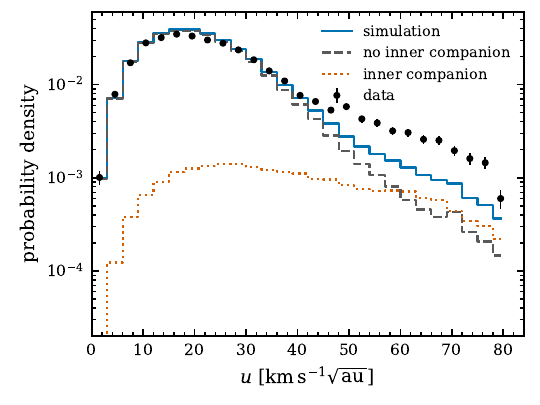}
\caption{Distribution of $u$ in the companion simulation (blue line) and in the data (points). The simulation uses the PARSEC MLR without a fit, and here it includes the measurement noise and both selection limits. Dashed and dotted lines are the realisations without and with a hidden companion.}
\label{fig:simcheck}
\end{figure}

A companion changes the measured $u$ through its mass, which adds to the total mass of the pair, and through its light. Its light blends with that of its host, so that the photocentre of the blend moves with the inner orbit \citep{vandeKamp1975}. We fit a straight line to this motion over the baseline of Gaia DR3. The slope of the line adds to the proper motion of the host. The residual increases the renormalised unit weight error (RUWE) of the Gaia astrometric fit \citep{Belokurov2020,Penoyre2020}.

We accept a realisation if its companions would remain hidden. A hidden companion is unresolved by Gaia and contributes less than $10\%$ of the $G$ flux. We assume that a brighter companion would change the XP spectrum. A hidden companion also leaves the RUWE below 3, a limit that we justify below.

The accepted companions change $u$ mainly through the photocentre. Their mass changes $u$ by at most $29\%$, whereas the photocentre motion adds a velocity comparable to the orbital velocity of the pair. In three quarters of the accepted realisations with a companion, $u$ changes by more than $20\%$. The companions that change $u$ by more than $20\%$ mostly have periods between about half a year and $1\,000$\,yr. At shorter periods the photocentre completes several orbits within the baseline, so its motion averages out in the proper motion. Companions with longer periods are mostly resolved or unstable.

The distribution $h_j$ is that of $u_0/\kappa$ in the accepted realisations with a companion. The simulation takes all masses from the PARSEC MLR, and for this scaling it uses the $\kappa$ that the two observed, blended magnitudes would give. The fraction $r_j$ is the share of accepted realisations that have a companion. Both are taken before the measurement noise and the selection limits, which the likelihood applies. We take $h_j$ and $r_j$ from the realisations of the pairs closest to pair $j$ in distance, separation and $\MG$ of the primary. The tail of $h_j$ is heavier for distant pairs (Fig.~\ref{fig:template_dist}), whose larger separations increase the added velocity through the factor $\sqrt{s}$.

The normalisation of the simulation is uncertain. With field-star companion fractions and no limit on RUWE, $3.5\%$ of the simulated pairs have a RUWE above 4, against $0.7\%$ of the observed pairs. Pairs with strong photocentre motion are thus largely absent from our sample. We apply no explicit cut on RUWE, but we require an astrometric fidelity of at least $0.5$ \citep{Rybizki2022}, which removes most sources with a poor astrometric fit. To reproduce this deficit, we reduced the companion fractions to a quarter and removed realisations with a RUWE above 3. The simulation then reproduces the core of the $u$ distribution and its decline at the upper limit (Fig.~\ref{fig:simcheck}).

The tail of the simulated distribution remains too low, with $6\%$ of the pairs at $u>45$ compared with $12\%$ in the data. The fractions $r_j$ of 3 to $10\%$ are also several times below the fitted mean of $0.29$. For this reason we fit the normalisation with $f_0$ and $\beta$. The simulation provides the distributions $h_j$ and the relative variation of $r_j$. With a fitted normalisation, the model reproduces the $u$ distributions of nearby and distant pairs separately (Sect.~\ref{sec:tests}). The simulation remains approximate in its model of the Gaia astrometry and its sharp detection thresholds, and it omits white-dwarf companions and chance alignments.

\section{Posterior computation and corner plots}\label{app:inference}
\subsection{Posterior computation}\label{sec:inference}
The posterior of the MLR parameters $\Theta$ and of the parameters of the outlier fraction is
\begin{equation}
 p(\Theta,f_0,\beta\mid\mathcal D)\propto\pi(\Theta,f_0,\beta)\prod_j\mathcal L_j(\Theta,f_0,\beta),
 \label{eq:posterior}
\end{equation}
with the likelihood of each pair from Eq.~(\ref{eq:marg}) and the priors $\pi$ of Appendix~\ref{app:model} (Table~\ref{tab:priors}). The metallicity enters only through the distributions $p_j(Z)$ of Eq.~(\ref{eq:zpost}), which are averages over $128$ posterior draws of the mixture fractions $\bm\omega$.

\begin{figure*}[t]
\centering
\includegraphics[width=0.8\textwidth]{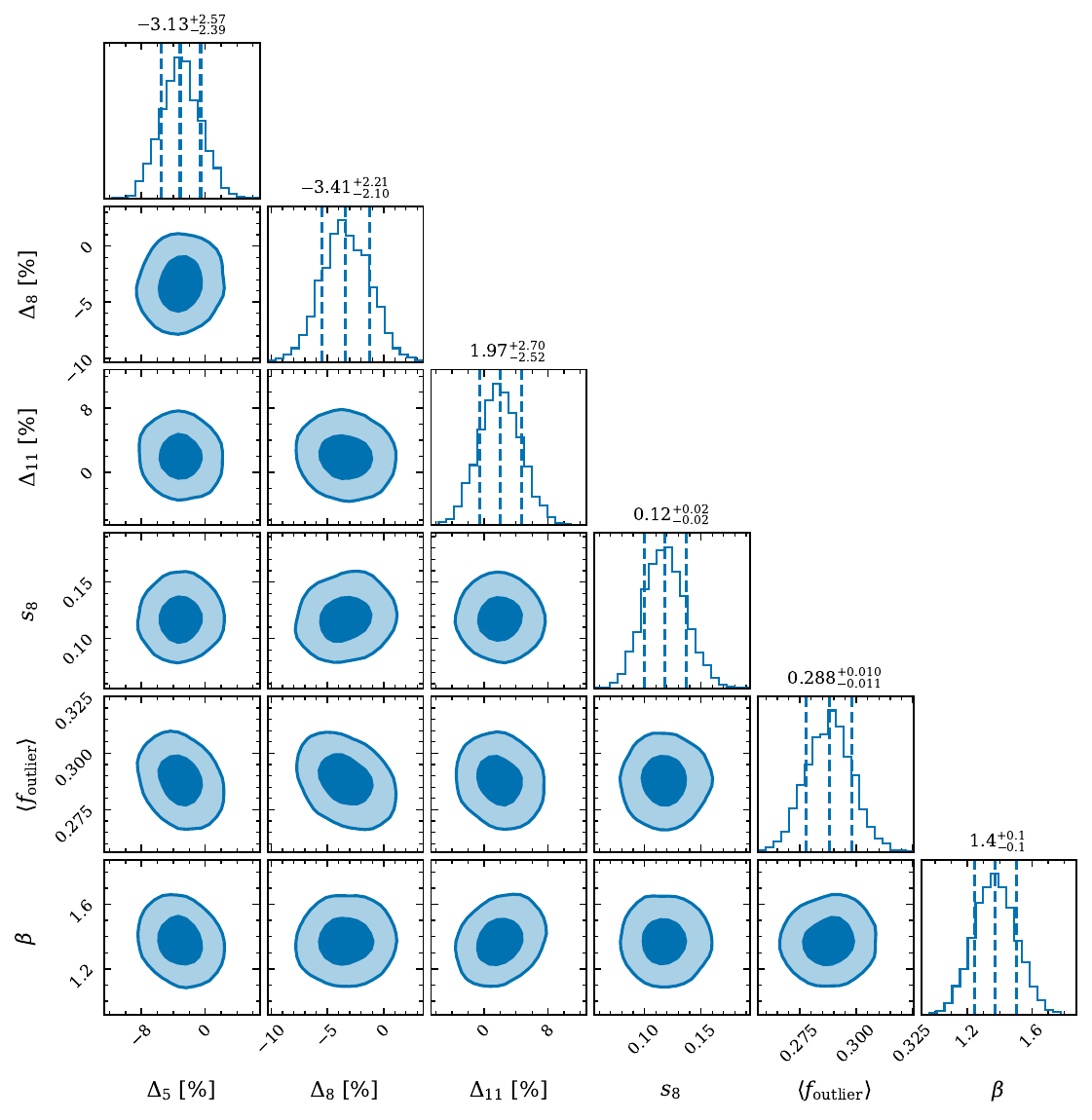}
\caption{Posterior of the adopted fit. $\Delta_{M_{\mathrm{G}}}$ is the percentage difference between the posterior mass and PARSEC at $Z=0$ and the given $M_{\mathrm{G}}$, $s_8$ is $\partial\log_{10}m/\partial Z$ at $M_{\mathrm{G}}=8$\,mag and $Z=0$, $\langle f_{\rm outlier}\rangle$ is the mean over the pairs of the outlier fraction $f_j$ of Eq.~(\ref{eq:fout}), and $\beta$ is its slope. $\langle f_{\rm outlier}\rangle$ refers to the model before the selection; its median of $0.288$ corresponds to a fraction of $0.33$ of the pairs in the selected sample, obtained from the posterior predictive distribution of Fig.~\ref{fig:upred}. Dashed lines mark the $16$, $50$ and $84\%$ quantiles, which are also given above each histogram as the median and the $68\%$ interval, and the contours enclose $39\%$ and $86\%$ of the samples.}
\label{fig:corner}
\end{figure*}

We implement the fit of $\bm\omega$ and the dynamical fit in NumPyro \citep{Phan2019NumPyro} on top of JAX \citep{Bradbury2018JAX} and run them on a single NVIDIA H100 GPU. The posteriors are sampled with the No-U-Turn Sampler \citep{HoffmanGelman2014NUTS} using four chains, each with $1\,000$ warm-up iterations followed by $7\,000$ retained draws and a target acceptance probability of $0.97$, from fixed random seeds.

The good component of Eq.~(\ref{eq:mix}) is evaluated for every pair before sampling on a geometric grid of $1\,024$ values of the mass scale $\kappa$ between $0.25$ and $2.5$, by Gauss--Legendre quadrature with $128$ nodes within $\pm14\,\sigma_{u,j}$ of $u_j$, and is interpolated in the logarithm during sampling. Replacing this grid by an adaptive quadrature changes the log-likelihood by less than $2\times10^{-5}$ for a test subsample.

The outlier component is tabulated in the same way, with the template $h_j$ of each pair, on a geometric grid of $400$ values of $\kappa$ over the same range, with $256$ nodes. The denominator of Eq.~(\ref{eq:sel}), the probability that a pair satisfies both selection limits, is tabulated on the same grids. For the Rice distribution this integral is given by the Marcum $Q$ function, $Q_1(a,t)=\Pr[\chi^2_2(a^2)>t^2]$, which we evaluate with the survival function of the non-central $\chi^2$ distribution.

\subsection{Corner plots}
Figure~\ref{fig:corner} shows the joint posterior of the quantities that matter for the interpretation of the MLR. The $32$ spline coordinates are not interpretable individually, so we show derived quantities at $Z=0$: the offset of the mass from PARSEC at $M_{\mathrm{G}}=5$, $8$ and $11$\,mag ($\Delta_5$, $\Delta_8$, $\Delta_{11}$), the metallicity slope $s_8=\partial\log_{10}m/\partial Z$ at $M_{\mathrm{G}}=8$\,mag, the mean outlier fraction $\langle f_j\rangle$ and the slope $\beta$ of Eq.~(\ref{eq:fout}). All marginals are close to Gaussian and unimodal. The correlations among the offsets and with the outlier fraction are discussed in Sects.~\ref{sec:res_mlr} and \ref{sec:res_fit}. In addition, $\beta$ is correlated with $\Delta_{11}$, with a coefficient of $0.26$. All fits, including the fit of $\bm\omega$ and the alternative fits of Appendix~\ref{sec:systematics}, converged, with $\hat R<1.01$ for every parameter and no divergent transitions. Here $\hat R$ is the split Gelman--Rubin statistic \citep{GelmanRubin1992,Vehtari2021}, which compares the variance between chains with the variance within chains and is close to $1$ for converged chains.

\section{Systematic checks}\label{sec:systematics}
This appendix gives the details of the tests summarised in Sect.~\ref{sec:tests}. Each test repeats the adopted fit with one element of the model changed, with the same metallicity posteriors $p_j(Z)$ and the same sampler settings. Table~\ref{tab:tests} lists the resulting changes of the median mass at $Z=0$. Here we also give the outlier fraction and the metallicity sensitivity of each fit, and we state how well it reproduces the observed distribution of $u$.

\subsection{Outlier model}\label{sec:sys_outlier}
\paragraph{Pairs with $u/\sigma_u>10$.} This test asks whether the pairs with small $u/\sigma_u$ bias the fit. These pairs have large $\sigma_u$ and high outlier fractions, so they depend most on the measurement kernel and the outlier model. Fitting only the $3\,862$ pairs with $u/\sigma_u>10$ raises the masses by 1 to $5\%$ at $\MG\le8$\,mag and changes them by about $1\%$ or less at $\MG\ge10$\,mag. The metallicity sensitivity exceeds that of PARSEC by $0.04$ to $0.05$, compared with $0.02$ to $0.03$ in the adopted fit (Sect.~\ref{sec:res_feh}).

The outlier fraction of this fit has a mean of $\langle f_j\rangle=0.19$, compared with $0.29$ in the adopted fit. It varies between the pairs as predicted by the companion simulation, with $\beta=1.01\pm0.23$. Both fits reproduce the distribution of $u$ of these pairs, as the cumulative distributions of model and data differ by less than the sampling noise at $95\%$ confidence. We take the difference between the two fits as the systematic uncertainty from the outlier and measurement models.

\paragraph{Pairs within $250$\,pc.} This test separates the effect of distance from that of $u/\sigma_u$. Nearby pairs have smaller $\sigma_u$, and their hidden companions are more often detected. Fitting only the $6\,076$ pairs with $d<250$\,pc raises the masses by up to $2.3\%$ at $\MG=6$ to 8\,mag, about half the change of the fit with $u/\sigma_u>10$. At $\MG\ge9$\,mag the masses change by at most $1\%$. The outlier fraction has a mean of $\langle f_j\rangle=0.21$, with $\beta=1.26\pm0.25$.

Within $250$\,pc, the fits with $u/\sigma_u>3$ and $u/\sigma_u>10$ differ by at most $1.4\%$ in mass at $Z=0$. The metallicity sensitivity of the second fit is nevertheless higher by up to $0.03$, more towards faint magnitudes. The difference in the mass level between the two thresholds thus comes mainly from the distant pairs with low $u/\sigma_u$. The difference in the metallicity sensitivity comes from pairs with low $u/\sigma_u$ at all distances.

\paragraph{One outlier distribution for all pairs.} This test replaces the outlier distributions $h_j$ of the individual pairs by one distribution pooled over all pairs, with one outlier fraction for all pairs. The masses are then higher by 2 to $8\%$ at $\MG\le6$\,mag and lower by 6 to $12\%$ at $\MG\ge8$\,mag. This fit does not describe how the distribution of $u$ changes with distance. When the pairs are split by distance, its cumulative distributions differ from those of the data by almost twice the sampling noise, whereas those of the adopted fit stay within it.

\subsection{Selection}\label{sec:sys_selection}
The two limits of Eq.~(\ref{eq:cut}) enter the likelihood through the selection correction of Eq.~(\ref{eq:sel}). We omit the correction for each limit in turn.

\paragraph{Upper limit.} Without the correction for the bound-orbit limit, the masses are higher by $0.4$ to $2.2\%$, most at $\MG=6$ to 8\,mag. The change is almost the same at all metallicities, so the metallicity sensitivity is unchanged.

\paragraph{Lower limit.} Without the correction for the lower limit, the masses are higher by up to $31\%$, most at $\MG=8$ to 10\,mag. At the bright end of the data, $\MG<4.3$\,mag, they are lower by up to $3\%$. The outlier fraction rises to $\langle f_j\rangle=0.41$, and its variation between the pairs, $\beta=3.5\pm0.2$, is far steeper than predicted by the companion simulation. The correction for the lower limit is therefore large, and the fitted relation depends on it.

\subsection{Solar mass anchor}\label{sec:sys_anchor}
\begin{figure}[t]
\centering
\includegraphics[width=\columnwidth]{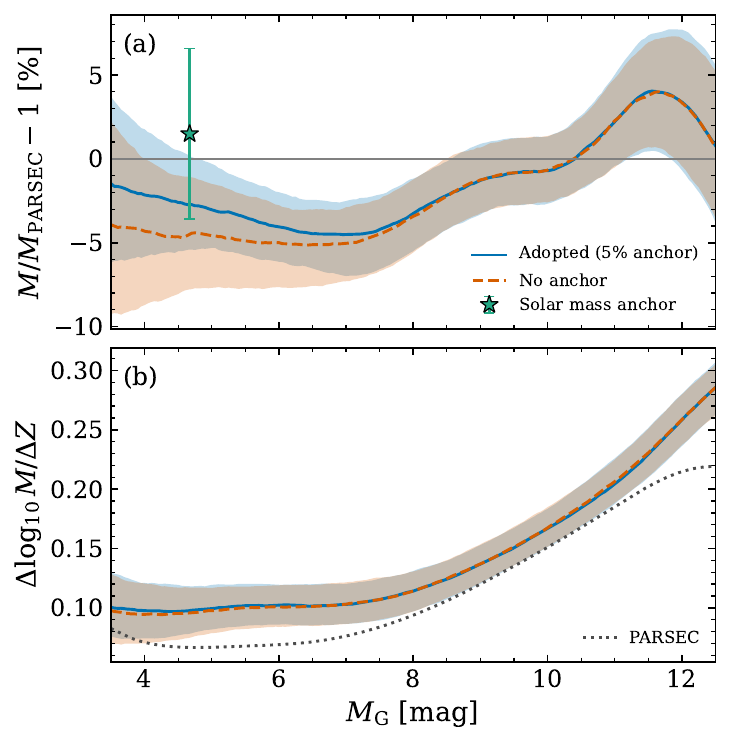}
\caption{Effect of the solar mass anchor. \emph{Panel (a):} mass at $Z=0$ relative to PARSEC at $Z=0$ for the adopted fit with the $5\%$ anchor (solid) and the same fit without the anchor (dashed); lines and bands are the posterior median and $68\%$ credible interval. The star marks the anchor with its $5\%$ width. \emph{Panel (b):} metallicity sensitivity $\Delta\log_{10}m/\Delta Z$ between $Z=-0.6$ and $+0.4$ for the two fits and for PARSEC (dotted).}
\label{fig:anchor}
\end{figure}

This test asks whether the solar mass anchor sets the mass scale of the MLR. We repeat the fit without the anchor, which we implement by widening it to $1$\,dex (Fig.~\ref{fig:anchor}). The masses at the bright end become slightly lower and their uncertainty slightly larger, but the change is smaller than the statistical uncertainty everywhere.

At $\MG\gtrsim8$\,mag the two fits are indistinguishable, and the metallicity sensitivity is the same at all magnitudes. The mass scale of the MLR is therefore set by the dynamics. The anchor only stabilises the bright end, where the pairs are fewer.

\subsection{Eccentricity distribution}\label{sec:sys_pu}
The shape of $p(\ut)$ (Sect.~\ref{sec:dynamics}) depends on the orbital elements of the population, and of these the eccentricity distribution is the least constrained. A more eccentric population spends more time near apocentre, at large separation and low speed. Its $p(\ut)$ is therefore shifted to lower values, and the masses inferred at a given $u$ are higher.

Changing the eccentricity index from the adopted $\alpha_e=1.23$ to $1.0$ or $1.5$ changes the masses by up to $2\%$. They are lower for $\alpha_e=1.0$ and higher for $\alpha_e=1.5$, by almost the same amount at all metallicities, so the metallicity sensitivity is not affected. The change is smaller than the statistical uncertainty of the MLR.

\end{appendix}

\end{document}